\documentclass[aps,prd,twocolumn,groupedaddress,10 pt]{revtex4-2}

\usepackage{graphicx}
\usepackage{dcolumn}
\usepackage{bm}
\usepackage{amsmath} 
\usepackage{amssymb}
\usepackage{hyperref}
\hypersetup{
	colorlinks=true,
	linkcolor=red,
	filecolor=magenta,      
	urlcolor=blue,
	pdftitle={Selective Thermalization},
	pdfpagemode=FullScreen,
	citecolor= blue,
}
\begin{document}


\title{Selective Thermalization, Chiral Excitations, and a Case of Quantum Hair
in the Presence of Event Horizons}


%
\author{Akhil U Nair}
	\email{p20200473@hyderabad.bits-pilani.ac.in}
\author{Rakesh K. Jha}
	\email{p20230070@hyderabad.bits-pilani.ac.in}
	\author{Prasant Samantray}
	\email{prasant.samantray@hyderabad.bits-pilani.ac.in}
	\author{Sashideep Gutti}
	\email{sashideep@hyderabad.bits-pilani.ac.in}
	\affiliation{Department of Physics,\\ \mbox{Birla Institute of Technology and Science, Pilani - Hyderabad Campus} \\Hyderabad, 500078, India}
	
	\date{\today}


\date{\today}

\begin{abstract}
The Unruh effect is a well-understood phenomenon,
where one considers a vacuum state of a quantum field in Minkowski spacetime, which appears to be thermally populated for a uniformly accelerating Rindler observer. In this article, we derive a variant of the Unruh effect involving two distinct accelerating observers and aim to address the following questions: (i) Is it possible to selectively thermalize a subset of momentum modes for the case of massless scalar fields, and (ii) Is it possible to excite only the left-handed massless fermions while keeping right-handed fermions in a vacuum state or vice versa? To this end, we consider a Rindler wedge $R_1$ constructed from a class of accelerating observers and another Rindler wedge $R_2$ (with $R_2 \subset R_1$) constructed from another class of accelerating observers such that the wedge $R_2$ is displaced along a null direction w.r.t $R_1$ by a parameter $\Delta$. By first considering a massless scalar field in the $R_1$ vacuum,  we show that if we choose the displacement $\Delta$ along one null direction, the positive momentum modes are thermalized, whereas negative momentum modes remain in vacuum (and vice versa if we choose the displacement along the other null direction). We then consider a massless fermionic field in a vacuum state in $R_1$ and show that the reduced state in $R_2$ is such that the left-handed fermions are excited and are thermal for large frequencies. In contrast, the right-handed fermions have negligible particle density and vice versa. We argue that the toy models involving shifted Rindler spacetime may provide insights into the particle excitation aspects of evolving horizons and the possibility of Rindler spacetime having a quantum strand of hair. Additionally, based on our work, we hypothesize that massless fermions underwent selective chiral excitations during the radiation-dominated era of cosmology.
\end{abstract}


\maketitle

\section{Introduction \label{Sec-1}}
Black holes are one of the most mysterious astrophysical objects that have ever been detected due to their classical and quantum properties. Black holes can emit particles and radiation due to quantum effects near the event horizon; this phenomenon is known in the literature as Hawking radiation \cite{Hawking.26.1344}. The spectrum of particles is shown to be thermal distribution at late times \cite{Hawking.26.1344}. The astrophysical black holes are formed due to the collapse of matter and radiation, forming an event horizon that prevents any object from escaping from a black hole to the outside.  
The black holes give rise to many unresolved questions, like the problem of the existence of spacetime singularities or information paradox. These are still unresolved and await concrete theoretical proofs and justifications.
\par Rindler spacetime has been used as a toy model to study and understand various aspects of black hole event horizon and Hawking radiation. This is because of the discovery of the Unruh effect \cite{Unruh-1976}. Though the Rindler spacetime is interesting in its own right, much of its importance comes from the fact that it happens to be near the horizon limit of generic black holes (non-extremal). Therefore, extensive work is being carried out in Rindler spacetime to eventually understand black hole horizons in the context of thermodynamics, entanglement entropy, and quantum information \citep{maldacena,witten1,bousso,suvrat,carlip}.
\par One area that is not yet fully understood is the quantum signatures of evolving horizons. If one considers that a black hole is formed by gravitational collapse, its mass evolves, and so does its event horizon (the boundary of the trapped region). The more relevant notion in this scenario is the apparent horizon. Based on the causal nature of the evolving horizon, it is classified broadly as a dynamical horizon if its tangent is spacelike (a typical collapsing of matter scenario) and a timelike tube if its tangent is timelike \cite{ashtekar2002} ( occurs in cosmological scenarios or during the collapse of homogeneous dust models like the Oppenheimer-Snyder collapse model \cite{oppie}). A uniform notion for evolving horizons is introduced in Ref.~\cite{Netta} known as holographic screens (the dynamical horizon then is a spacelike future holographic screen). There is an even less explored scenario where the evolving horizon is null. This scenario occurs during the radiation-dominated era in cosmological evolution \cite{bendov,raviteja}. The work done in this article might prove useful for deducing the clues towards possible quantum signatures for this scenario, as argued in the paragraphs below. 
\par
\begin{figure}[h]
	\centering
	\includegraphics[scale=0.8]{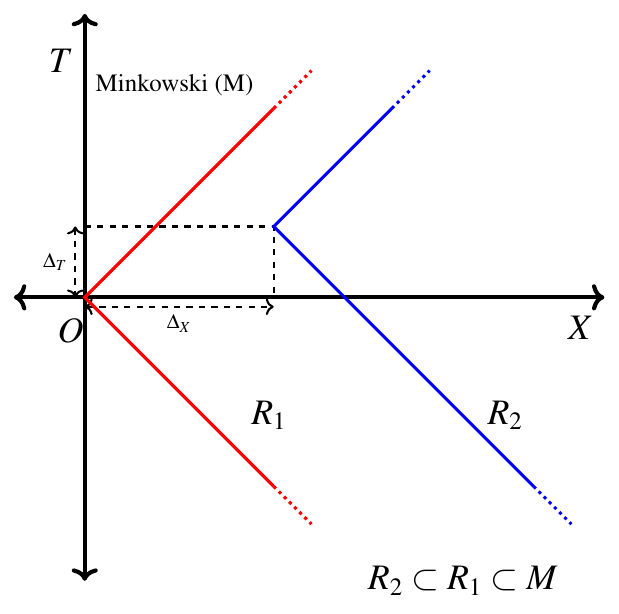}
	\caption{Spacelike-shifted Rindler spacetimes  ($R_1$) and ($R_2$). Here $\Delta_x$ and $\Delta_t$ represents the shift in $X$ and $T$ coordinate axis.
	\label{Fig:0}}
\end{figure}

 The Rindler spacetimes provide interesting toy models for understanding the quantum signatures of evolving horizons. When one constructs two different Rindler spacetimes using two separate classes of uniformly accelerating observers (say $R_1$ and $R_2$ respectively), we can arrange the observers in such a way that $R_2$ is a subsystem of $R_1$ ( $R_2 \subset R_1$). The Fig.~\ref{Fig:0} illustrates the situation. For $R_2$ to be a subsystem of $R_1$, it is easy to see that the bifurcation point of $R_2$ is spacelike or null separated from $R_1$ as shown in Figs.~\ref{Fig:0}, \ref{Fig:1}, and \ref{Fig:2}.
\par A good toy model for the case of dynamical horizons is a nested sequence of spacelike shifted Rindler spacetimes, as argued in Ref.~\cite{Gutti2023}. This is because a dynamical horizon in the context of black holes makes a series of spacelike jumps in the increasing radial direction. The situation resembles a nested sequence of Rindler wedges. The nested sequence of wedges that occur in the case of dynamical horizons (or future holographic screens)  has been used to prove the important results concerning the increase in entanglement entropy in the outer region of a black hole in ADS/CFT context \cite{Netta-Wall}.  A review of relevant literature on entanglement in quantum field theoretic setting is discussed in \citep{witten1,tatsuma,headrik,casini,cardy, Hollands2017}.
\par Few papers have dealt with shifted Rindler scenarios and have interesting results. In Ref.~\cite{kinjalkpaddy}, the authors consider Rindler spacetime $R_1$ and another Rindler spacetime $R_2$ shifted along the spatial direction. They showed that if quantum fields in $R_1$ are in vacuum, $R_2$ has a thermal spectrum of particles. In Ref.~\cite{Gutti2023}, the authors addressed a question as to the uniqueness of particle content in $R_2$ starting from a vacuum state in Minkowski spacetime depending on whether we consider $R_1$ between $R_2$ and Minkowski spacetime or not ($R_2 \subset R_1 \subset M$ with M as Minkowski spacetime).
\par Here are a few interesting questions, some of which are answered in this article. Suppose there is a spacetime with a quantum field living on it in a vacuum state. Can we construct subsystems of this system demarcated due to horizons to partially excite or thermalize a subset of modes? Can Rindler spacetime exhibit a strand of hair? We show that the answers to the above questions are affirmative. We choose $R_1$ as a system with a massless scalar field or a massless dirac field in a vacuum state. We then define its subsystem as $R_2$ shifted along one of the null directions (see Figs.~\ref{Fig:1}, \ref{Fig:2}) and then find out the state of the scalar or Dirac fields. We obtain that, depending on the direction of the null shift, positive momentum modes (negative momentum modes) are in a thermal state. In contrast, the negative momentum modes (positive momentum modes) remain in a vacuum.
\par When massless Dirac fields are considered, though there is no coupling between left-handed and right-handed fermions in Minkowski spacetime, there is a non-trivial coupling between them in Rindler spacetimes. We show that the left-handed (right-handed) fermions are in an excited state (arguably thermal), whereas the right-handed (left-handed) fermions have negligible excitation (despite the coupling between them in Rindler spacetime). We, therefore, can produce Chiral excitations using Rindler horizons.
\par When we consider gravitational situations where the apparent horizons evolve along the null direction during the radiation-dominated era \cite{bendov, raviteja}, the results in this article suggest that there might be chiral excitations of massless fermions. Whether this holds true or not, and if it is true, are there any important consequences that have not yet been answered.
\par The results in the article show that given a Rindler spacetime with a particle content with selective thermalization, it is possible to predict the causal positioning of the superset of the Rindler spacetime (purification of the Rindler spacetime). This gives quantum hair to Rindler spacetime.
\par In the following sections, we will provide a comprehensive view of how selective thermalization exists and manifests in our scenario. In Sec.~\ref{Sec-2}, we briefly discuss the setup which leads to selective thermalization. We will introduce the relevant coordinate systems and their mutual relations. In Sec.~\ref{Sec-3}, we present detailed methods and techniques on how selective thermalization manifests for the case of Scalar fields. The first part of Sec.~\ref{Sec-4} introduces the Dirac fields in our setup, and it reproduces some of the existing results, such as the Unruh effect for the Fermionic fields. In the subsequent parts of Sec.~\ref{Sec-4}, we provide the detailed theoretical framework underlying our approach on how selective Chiral excitations manifest for massless Dirac fields. Finally, in Sec.~\ref{Sec-5}, we summarize our conclusions and discuss the implications of our results along with the potential future direction for further research on this problem and associated problems.

\section{The SetUp \label{Sec-2}}
\subsection{Rindler coordinates \label{Subsec-2.1}}

In this article, we investigate the possibility of selective excitations of fields wherein we use classes of accelerating observers and arrange them so that only a subset of the modes are excited while the other modes remain in a vacuum state. We demonstrate this by analyzing two cases: one involving a free massless scalar field and the other using a free massless Dirac field. The spacetimes under consideration are Minkowski and Rindler, with coordinates denoted by $\left( X, T \right)$ for Minkowski spacetime M and $\left(x_i, t_i\right)$ for the various Rindler frames $R_i$, respectively. When one considers two Rindler frames as in Fig~\ref{Fig:0}, the goal is that $R_2 \subset R_1$, so we can choose $R_1$ such that the bifurcation point of $R_1$ is at the origin of $M$, this is illustrated in Fig.~\ref{Fig:0}. The scenario where $R_2\subset R_1$ is possible only when the bifurcation point of $R_2$ is separated from the bifurcation point of $R_1$ by a spacelike interval or a null interval. The separation by spacelike intervals and a few issues have been explored in \cite{LochanPady2021, Gutti2023}. We note that upon a spacelike shift, one can carry out a Lorentz Boost and bring $R_2$ such that the bifurcation point lies on the X-axis [see in Fig.~\ref{Fig:0}]. In the references \cite{Gutti2023} and \cite{LochanPady2021}, it was shown that when assuming $R_1$ to be in a vacuum, the state in $R_2$ is thermal (at least in the large frequency limit). The result, therefore, carries over to a shift along an arbitrary spacelike interval (with a restriction that $R_2$ is a subset of $R_1$) because the Fig.~\ref{Fig:0} can be transformed using a boost into a shift along the X-axis in $M$. The non-trivial scenario is when we choose the shift along the null direction.  There are two scenarios we consider in this article; one is a shift of Rindler-2 ($R_2$) frame in the ``$V-\text{axis}$", and the other is a shift in the ``$U-\text{axis}$," both are in the null direction  (see Figs.~\ref{Fig:1} and \ref{Fig:2} respectively).
\subsubsection{Null-Shift in $V-$axis:}
We label the coordinate of the Rindler-1 ($R_1$) frame as $\left(x_1,t_1\right)$ and that of Rindler-2 ($R_2$) as $\left(x_2,t_2\right)$. 
\begin{figure}[ht]
	\centering
	\includegraphics[scale=0.7]{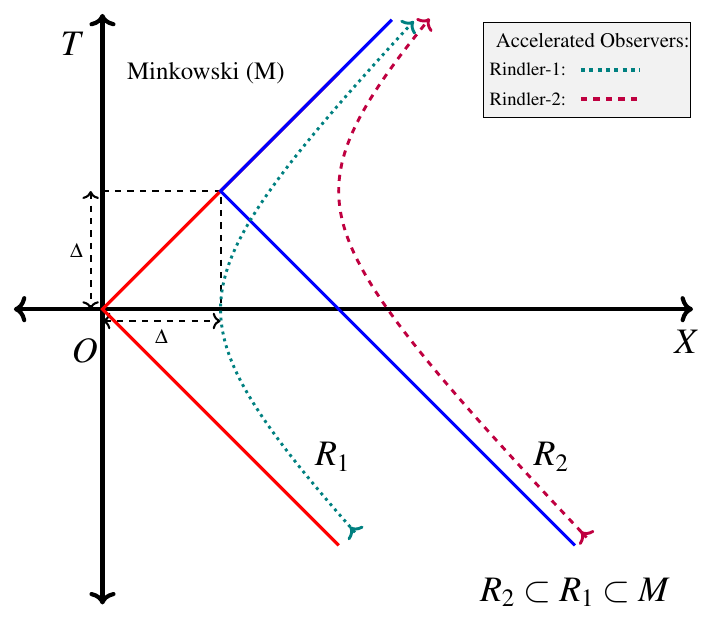}
	\caption{Null-shifted Rindler spacetimes  ($R_1$) and ($R_2$) along the $V-$axis.\label{Fig:1}}
\end{figure}
The relationship between $R_1$ coordinates and the Minkowski ($M$) coordinates, as illustrated in Fig.~\ref{Fig:1} can be expressed as follows:
\begin{equation}
	T = \frac{e^{g_1 x_1}}{g_1} \sinh(g_1 t_1), \label{Eq:2.1.1.1}
\end{equation}
\begin{equation}
	X = \frac{e^{g_1 x_1}}{g_1} \cosh(g_1 t_1) .\label{Eq:2.1.1.2}
\end{equation}
In terms of the Light-cone coordinates, this can be written as,
\begin{equation}
    U_M = T-X = -\frac{e^{-g_1 u_1}}{g_1}, 
    \label{Eq:2.1.1.3}
\end{equation}
\begin{equation}
    V_M = T+X = \frac{e^{g_1 v_1}}{g_1}. 
    \label{Eq:2.1.1.4}
\end{equation}
Likewise, we define the $R_2$ frame with the coordinates $\left(x_2, t_2\right)$, such that it is null-shifted in the $V$-axis. It should be noted that the $V$-axis is given by $U=0$, which implies $T=X$. The relationship of $R_2$ and that of the Minkowski ($M$) coordinates are given as:
\begin{equation}
    T = \frac{e^{g_2 x_2}}{g_2} \sinh(g_2 t_2)+ \Delta, 
    \label{Eq:2.1.1.5}
\end{equation}
\begin{equation}
    X = \frac{e^{g_2 x_2}}{g_2} \cosh(g_2 t_2)+ \Delta, 
    \label{Eq:2.1.1.6}
\end{equation}
where $\Delta$ represents the shift in the coordinate axis.
Now, expressed in light-cone coordinates, we have
\begin{equation}
	U_M = T-X = -\frac{e^{-g_2 u_2}}{g_2}, \label{Eq:2.1.1.7} 
\end{equation}
\begin{equation}
    V_M = T+X = \frac{e^{g_2 v_2}}{g_2} + 2\Delta .
    \label{Eq:2.1.1.8}
\end{equation}
Note, below, we state the relation between the coordinates of $R_i$ and the corresponding light-cone coordinates of $R_i$ as,
\begin{eqnarray}
    u_i &=& t_i-x_i, \label{Eq:2.1.1.8.a}\\
    v_i &=& t_i+x_i,\label{Eq:2.1.1.8.b}
\end{eqnarray}
where $i \in (1,2)$ labels the Rindler frame $R_1$ or $R_2$. The general form of the metric in any of the Rindler frames has the form:
\begin{equation}
	ds^2 = e^{2 g_i x_i}\big(dt^2_i-dx^2_i\big), \label{Eq:2.1.1.9} 
\end{equation}
here also, $i \in (0,1)$, labels a particular Rindler frame under consideration. Note that, for the rest of the calculations, we consider the acceleration parameter for different Rindler frames, $g_1 = g_2 =g$; this simplifies the problem.
To have an understanding and interpretation of the results of our article in terms of accelerating observers, we identify a canonical trajectory of an observer in $R_1$ in a ``teal dotted" hyperbola in Fig.~\ref{Fig:1}. Similarly, we plot a canonical accelerating trajectory for the frame $R_2$ using a ``purple dashed" hyperbola.  The accelerating observers in $R_1$ have the information from accelerating observers in $R_2$ during the entire trajectory. In contrast, accelerating observers in frame $R_2$ will see the accelerating observers in $R_1$ disappearing behind the horizon and, therefore, have access to observers in $R_1$ for a restricted part of their trajectory. 

\subsubsection{Null-Shift in $U$-axis:}
\begin{figure}[ht]
	\centering
	\includegraphics[scale=0.7]{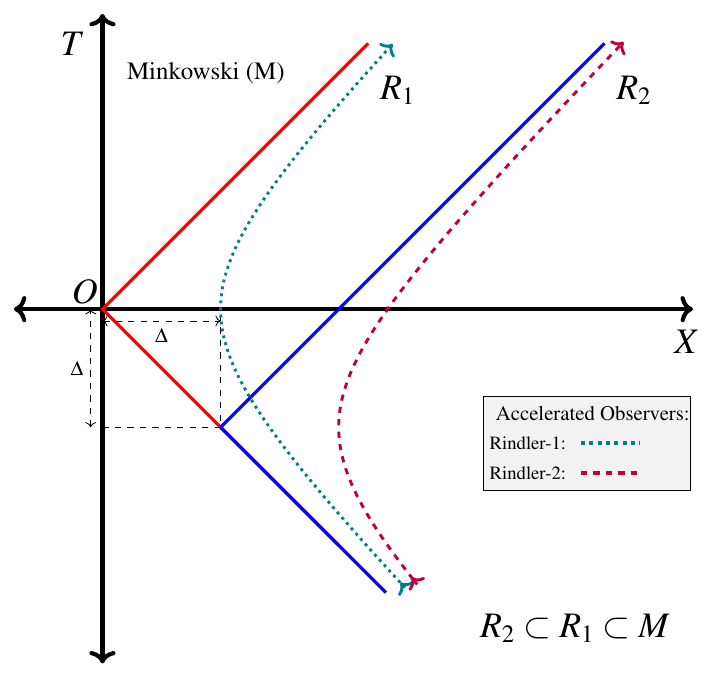}
	\caption{Null-shifted Rindler spacetimes  ($R_1$) and ($R_2$) along the $U-$axis.\label{Fig:2}}
\end{figure}

Similarly, coming to the other scenario as seen in Fig.~\ref{Fig:2}, where we have considered a null-shift of $R_2$ in the $U-$axis, i.e., the $V=0$ case. The relationship between the $R_2$ frame with coordinates $\left(x_2, t_2\right)$ and the Minkowski coordinates are given as:
\begin{eqnarray}
	T &=& \frac{e^{g_2 x_2}}{g_2} \sinh(g_2 t_2)- \Delta, \label{Eq:2.1.2.1} \\
	X &=& \frac{e^{g_2 x_2}}{g_2} \cosh(g_2 t_2)+ \Delta. \label{Eq:2.1.2.2}
\end{eqnarray}
Thus, expressed in light-cone coordinates, we get,
\begin{eqnarray}
	U_M &=& T-X = -\frac{e^{-g_2 u_2}}{g_2}- 2\Delta, \label{Eq:2.1.2.3} \\
	V_M &=& T+X = \frac{e^{g_2 v_2}}{g_2} .\label{Eq:2.1.2.4}
\end{eqnarray}
Note that, as already mentioned, the form of the Rindler metric remains the same as in Eq.~(\ref{Eq:2.1.1.9}).
To carry out an analysis using accelerating observers, 
We now identify a canonical trajectory of an observer in $R_1$ in a ``teal dotted" hyperbola in Fig.~\ref{Fig:1}. We similarly plot a canonical accelerating trajectory for the frame $R_2$ using a ``purple dashed" hyperbola. The accelerating observers in $R_2$ have the information from accelerating observers in $R_1$ during the entire trajectory. In contrast, accelerating observers in frame $R_1$ start obtaining information from observers in $R_2$ only after they cross the horizon in $R_2$.

\section{Selective Excitation for massless Scalar Fields \label{Sec-3} }
This section considers a more straightforward case of a free massless scalar field in a vacuum state in the frame $R_1$. We will calculate the particle content in $R_2$ for the case where $R_2$ is shifted from $R_1$ along the $V$-axis. It is shown below that the particle content of the left-moving modes gets excited, which matches with the thermal spectrum for large frequencies. The right-moving modes, however, remain in a vacuum. Based on symmetry considerations, we similarly predict the particle content where $R_2$ is shifted along the $U$-axis. 

\subsection{Mode expansion in Rindler spacetime \label{Subsec-3.1}}
Consider a free massless scalar field in $(1+1)$ dimensional Rindler spacetime. The equations of motion are given by the Klein-Gordon Equation,
\begin{equation}
	\Box \;\hat{\phi}(x^\mu) = 0.\label{Eq:3.1.0.1}
\end{equation}
The mode expansion of the field operator $\hat{\phi}(x^\mu)$ is similar for any of the Rindler frames and is given by,
\begin{equation}
    \begin{split}
	\hat{\phi}(x_i,t_i) =
	\int_{0}^{\infty} dp \bigg[&\hat{c}_i(p) h_i(p) + 
        \hat{c}_i^\dagger(p)h^*_i(p) \\ &+  \hat{d}_i(p)j_i(p) + \hat{d}_i^\dagger(p)j^*_i(p)  \bigg], 
    \end{split}
    \label{Eq:3.1.0.2}
\end{equation}
where $h$ and $j$ are the right-moving and left-moving modes, respectively and the $i$, labels the Rindler frame. For a given Rindler frame, the modes can be represented by plane waves and are given as,
\begin{equation}
	h_i(p) = \frac{e^{i\left( -\omega_p t_i + p x_i \right)}}{\sqrt{4 \pi \omega_p}} \label{Eq:3.1.0.3}
\end{equation} 
and 
\begin{equation}
	j_i(p) = \frac{e^{i\left( -\omega_p t_i - p x_i \right)}}{\sqrt{4 \pi \omega_p}}. \label{Eq:3.1.0.4}
\end{equation}
These plane wave solutions can be normalized under the Klein-Gordon inner product and can be shown to form a complete basis for describing the field operator. We note that the analysis presented below on free massless scalar field uses Wald's \cite{Wald1993} representation of the Klein-Gordon inner product and is given as,
\begin{equation}
	\langle \phi_1,\phi_2 \rangle = -i \int_{\Sigma} \sqrt{-\sigma}d\Sigma^{\mu} \left( \phi_1^* \partial_\mu \phi_2 - \phi_2\partial_\mu \phi_1^* \right),
	\label{Eq:3.1.0.5}
\end{equation}
where $\Sigma$ is an appropriate constant time Cauchy hypersurface for the particular Klein-Gordon equation.
Additionally, we note that for a massless scalar field, the on-shell condition gives $\omega_p=p>0$, i.e., $p$ is non-negative; thus, the Eq.~(\ref{Eq:3.1.0.2}) for the case of $R_1$ can be rewritten using Eq.~(\ref{Eq:3.1.0.3}) and Eq.~(\ref{Eq:3.1.0.4}) as,
\begin{equation}
    \begin{split}
    \hat{\phi}(x_1,t_1) =& \int_{0}^{\infty} dp        \bigg[\hat{c}_1(p)\frac{e^{-ip\left( t_1 -  x_1 \right)}}{\sqrt{4 \pi p}} +  \hat{c}_1^\dagger(p)\frac{e^{i p\left( t_1 -  x_1 \right)}}{\sqrt{4 \pi p}}\\
    &+ \hat{d}_1(p) \frac{e^{-i p\left( t_1 +  x_1 \right)}}{\sqrt{4 \pi p}}+ \hat{d}_1^\dagger(p)\frac{e^{i p \left( t_1 +  x_1 \right)}}{\sqrt{4 \pi p}}  \bigg]. 
    \end{split}
    \label{Eq:3.1.0.6}
\end{equation}
The operators $\hat{c}_1^\dagger(p),\hat{d}_1^\dagger(p)$ and $\hat{c}_1(p),\hat{d}_1(p)$ are the usual creation and annihilation operators of the field and they obey the standard commutation relations:
\begin{equation}
    \begin{split}
        [\hat{c}_1(p),\hat{c}_1^\dagger(p')] =& \delta(p-p'),\\
        [\hat{c}_1(p),\hat{c}_1(p')] =& 0,\\
        [\hat{c}_1^\dagger(p),\hat{c}_1^\dagger(p')] =& 0.\\
    \end{split}
    \label{Eq:3.1.0.7}
\end{equation}
Similarly, the mode expansion of the field operator in $R_2$ is,
\begin{equation}
    \begin{split}
        \hat{\phi}(x_2,t_2) =& \int_{0}^{\infty} dp \bigg[\hat{c}_2(p)\frac{e^{-i p\left( t_2 -  x_2 \right)}}{\sqrt{4 \pi p}} +  \hat{c}_2^\dagger(p)\frac{e^{i p\left( t_2 -  x_2 \right)}}{\sqrt{4 \pi p}}\\
        &+\hat{d}_2(p) \frac{e^{-i p \left( t_2 +  x_2 \right)}}{\sqrt{4 \pi p}}+ \hat{d}_2^\dagger(p)\frac{e^{i p \left( t_2 +  x_2 \right)}}{\sqrt{4 \pi p}}  \bigg]. 
    \end{split}
    \label{Eq:3.1.0.8}
\end{equation}
The field operator $\hat{\phi}(x_i,t_i) $, as evident from the above expressions [Eq.~(\ref{Eq:3.1.0.7}), Eq.~(\ref{Eq:3.1.0.8})], can be written as,
\begin{equation}
    \hat{\phi}(x_i,t_i)=\hat{\phi}(U)+\hat{\phi}(V),
    \label{Eq:3.1.0.9}
\end{equation}
thus implying that the left-moving and right-moving sectors are decoupled.
\subsection{Particle content of the Right moving modes:\label{Subsec-3.2}}
Since the right moving and left moving modes are decoupled, they form a complete basis independently \citep{Crispino}. Now, for the following analysis, it is more convenient to express the modes in terms of null coordinates $u_i,v_i$. We note the following relations, i.e., Eq.~(\ref{Eq:2.1.1.3}), Eq.~(\ref{Eq:2.1.1.4}), Eq.~(\ref{Eq:2.1.1.7}), and Eq.~(\ref{Eq:2.1.1.8}) for our calculations. We estimate the particle content by evaluating the Bogoliubov coefficients. We note  that the modes of $R_2$ are related to the modes of $R_1$ through a Bogoliubov transformation of the following form:
\begin{equation}
     h_{2}(q) = \int_{0}^{\infty} dk \bigg(\alpha_{21u}(q,k)h_{1}(k) + \beta_{21u}(q,k)h_{1}^*(k)\bigg),  \label{Eq:3.2.0.1}
\end{equation}
\begin{equation}
	h_{1}(k) = \int_{0}^{\infty} dq \bigg(\alpha^*_{21u}(q,k)h_{2}(q) - \beta_{21u}(q,k)h_{2}^*(q)\bigg).
	\label{Eq:3.2.0.2}
\end{equation}
Note that the subscript ``$u$" in the Bogoliubov coefficients labels that it is for the right moving modes  [Eg, $\alpha_{21u}(q,k)$].
Based on the above relations [see Eq~(\ref{Eq:3.2.0.1}) and Eq~(\ref{Eq:3.2.0.2})], we can write the creation and annihilation operators of $R_2$  in terms of $R_1$ creation and annihilation operators as \cite{mukhanov_winitzki_2007},
\begin{equation}
      \hat{c}_2(p) = \int_{0}^{\infty} dk \bigg(\alpha^*_{21u}(p,k)\hat{c}_1(k) - \beta^*_{21u}(p,k)\hat{c}_1^\dagger(k)\bigg), \label{Eq:3.2.0.3}
\end{equation}
\begin{equation}
       \hat{c}_2^\dagger(p) = \int_{0}^{\infty} dk \bigg(\alpha_{21u}(p,k)\hat{c}_1^\dagger(k) - \beta_{21u}(p,k)\hat{c}_1(k)\bigg).
	\label{Eq:3.2.0.4}
\end{equation}
Now, coming back to the field operators in $R_1$ and $R_2$, it can be asserted that at any point  $R_2$, i.e., in the domain of the spacetime where both coordinates frames overlap \cite{mukhanov_winitzki_2007}, the field operator $\hat{\phi}(U)$ can be independently expressed using the right-moving modes of $R_1$ or right moving modes of $R_2$ as a basis \cite{Valdivia-Mera.200109869V}, i.e.,
\begin{equation}
      \begin{split}
      	\phi(U) =&\int_{0}^{\infty} dk  \bigg[\hat{c}_1(k)\frac{e^{-i k u_1}}{\sqrt{4 \pi k}} + \hat{c}_1^\dagger(k)\frac{e^{i k u_1}}{\sqrt{4 \pi k}}\bigg] \\
      	&=\int_{0}^{\infty} dp \bigg[\quad\hat{c}_2(p)\frac{e^{-i p u_2}}{\sqrt{4 \pi p}} +  \hat{c}_2^\dagger(p)\frac{e^{i p u_2}}{\sqrt{4 \pi p}}\bigg],
      	\label{Eq:3.2.0.5}
      \end{split}
\end{equation}
where we have used the relations Eq.~(\ref{Eq:2.1.1.8.a}) and Eq.~(\ref{Eq:2.1.1.8.b}) to write the mode expansions in lightcone coordinates of $R_1$ and $R_2$. The relation between the null coordinates of $R_1$ and $R_2$, can be deduced from  Eq.~(\ref{Eq:2.1.1.3}) and Eq.~(\ref{Eq:2.1.1.7}) , 
\begin{equation}
	\frac{e^{-g u_1}}{g} = \frac{e^{-g u_2}}{g},
	\label{Eq:3.2.0.6}
\end{equation}
which implies $u_1=u_2$. From this, it is clear that LHS and RHS of Eq.~(\ref{Eq:3.2.0.5}) are, therefore, functions of $u_2$. Now, to evaluate the Bogoliubov coefficients, we multiply both sides of  Eq.~(\ref{Eq:3.2.0.5}) with,
\begin{equation}
    \begin{split}
    	&\int_{-\infty}^{\infty}\frac{du_2}{\sqrt{2\pi}}e^{i\Omega u_2}\int_{0}^{\infty} \frac{dk}{\sqrt{2\pi}} \bigg[\hat{c}_1(k)\frac{e^{-iku_1}}{\sqrt{2 k}} +  \hat{c}_1^\dagger(k)\frac{e^{iku_1}}{\sqrt{2 k}}\bigg] \\ &=\int_{-\infty}^{\infty}\frac{du_2}{\sqrt{2\pi}}e^{i\Omega u_2}\int_{0}^{\infty} \frac{dp}{\sqrt{2\pi}} \bigg[\hat{c}_2(p)\frac{e^{-ipu_2}}{\sqrt{2 p}} + \hat{c}_2^\dagger(p)\frac{e^{ipu_2}}{\sqrt{2 p}}\bigg].
    	\label{Eq:3.2.0.7}
    \end{split}
\end{equation}
Further simplification yields,
\begin{equation}
     \begin{split}
     	\hat{c}_2(\Omega) = \int_{0}^{\infty}&dk \bigg[\hat{c}_1(k)\sqrt{\frac{ \Omega}{k}}\int_{-\infty}^{\infty}\frac{du_2}{2\pi}e^{i(\Omega u_2-ku_1)} \\ 
     	&+ \hat{c}_1^\dagger(k)\sqrt{\frac{ \Omega}{k}}\int_{-\infty}^{\infty}\frac{du_2}{2\pi}e^{i(\Omega u_2+ku_1)}\bigg]. 
     	\label{Eq:3.2.0.8}
     \end{split}
\end{equation}
A comparison of the Eq.~(\ref{Eq:3.2.0.8}) with Eq.~(\ref{Eq:3.2.0.3}) yields the expressions for the Bogoliubov coefficients for the right moving modes as
\begin{equation}
     \alpha^*_{21u}(k,\Omega) =  \sqrt{\frac{ \Omega}{k}}\int_{-\infty}^{\infty}\frac{du_2}{2\pi}e^{i(\Omega u_2-ku_1)},
      \label{Eq:3.2.0.9}
\end{equation}
and, 
\begin{equation}
    \beta^*_{21u}(k,\Omega) =  -\sqrt{\frac{ \Omega}{k}}\int_{-\infty}^{\infty}\frac{du_2}{2\pi}e^{i(\Omega u_2+ku_1)}.
    \label{Eq:3.2.0.10}
\end{equation}
Using $u_1 = u_2$ and evaluating the integrals in the Eq.~(\ref{Eq:3.2.0.9}) and Eq.~(\ref{Eq:3.2.0.10}), we obtain the results for the Bogoliubov coefficients as
\begin{equation}
     \quad\alpha^*_{21u}(k,\Omega) = \sqrt{\frac{ \Omega}{k}} \delta(\Omega-k), \qquad \forall \:\Omega=k, 
     \label{Eq:3.2.0.11}
\end{equation}
and
\begin{equation}
	\beta^*_{21u}(k,\Omega) = -\sqrt{\frac{ \Omega}{k}} \delta(\Omega+k) = 0, \qquad \forall\: \Omega>0.
	\label{Eq:3.2.0.12}
\end{equation}
Note that, since both $\Omega$ and $k$ are positive quantities, it can be concluded that $ \beta^*_{21u}(k,\Omega)= 0, \quad (\forall\: \Omega>0 \:\&\: \Omega=k)$. The fact that $\beta_{21u}(k,\Omega)=0$, implies that the $R_2$ modes ($h_{2}$) are made of purely positive frequency functions. Thus, there is no mixing of modes manifesting in the context of right-moving modes, where mixing of modes is considered as a crucial criterion for having non-trivial ground states and, thus, non-zero particle distribution \cite{Padmanabhan_2010}. Therefore, starting from the vacuum in $R_1$, the right-moving modes remain in a vacuum state since the particle number density for the right-moving modes is zero.

\subsection{Thermalization of Left moving modes:\label{Subsec-3.3}}
The above line of analysis can be repeated for the case of left-moving modes. We show in this section that, unlike the right-moving modes that remain in a vacuum, the left-moving modes have a thermal distribution of particles. To prove this, we note that in this case also,  the Bogoliubov transformation between the left-moving modes of $R_2$ and $R_1$ can be defined as, 
\begin{equation}
    j_2(q) = \int_{0}^{\infty} d k \bigg(\alpha_{21v}(q,k)j_1(k) + \beta_{21v}(q,k)j_1^*(k)\bigg),  
    \label{Eq:3.3.0.1}
\end{equation}
\begin{equation}
	j_1(k) = \int_{0}^{\infty} d q \bigg(\alpha^*_{21v}(q,k)j_2(q) - \beta_{21v}(q,k)j_2^*(q)\bigg),
	\label{Eq:3.3.0.2}
\end{equation}
where, $j_2(q)$ and $j_1(k)$ are the positive frequency modes in $R_2$ and $R_1$ respectively. 
Proceeding further gives the operator relationship between the $R_2$ creation and annihilation operators as a linear combination of the creation and annihilation operators of $R_1$, i.e.,
\begin{equation}
     \hat{d}_2(p) = \int_{0}^{\infty} d k \bigg(\alpha^*_{21u}(p,k)\hat{d}_1(k) - \beta^*_{21u}(p,k)\hat{d}_1^\dagger(k)\bigg), \label{Eq:3.3.0.3}
\end{equation}
\begin{equation}
	\hat{d}_2^\dagger(p) = \int_{0}^{\infty} dk \bigg(\alpha_{21u}(p,k)\hat{d}_1^\dagger(k) - \beta_{21u}(p,k)\hat{d}_1(k)\bigg).
	\label{Eq:3.3.0.4}
\end{equation}
 Using the null coordinates Eq.~(\ref{Eq:2.1.1.8.a}) and Eq.~(\ref{Eq:2.1.1.8.b}) to rewrite the field operators in terms of both the basis as,
\begin{equation}
     \begin{split}
     	\phi(V) = \int_{0}^{\infty}& \frac{dk}{\sqrt{2\pi}} \bigg[\hat{d}_1(k)\frac{e^{-ikv_1}}{\sqrt{2 \omega_k}} +  \hat{d}_1^\dagger(k)\frac{e^{ikv_1}}{\sqrt{2 \omega_k}}\bigg]\\
     	&= \int_{0}^{\infty} \frac{dp}{\sqrt{2\pi}} \bigg[\hat{d}_2(p)\frac{e^{-ipv_2}}{\sqrt{2 \omega_p}} + \hat{d}_2^\dagger(p)\frac{e^{ipv_2}}{\sqrt{2 \omega_p}}\bigg].
     	\label{Eq:3.3.0.5}
     \end{split}
\end{equation} 
From Eq.~(\ref{Eq:2.1.1.4}) and Eq.~(\ref{Eq:2.1.1.8}), the relation between the null coordinates of $R_1$ and $R_2$ can be computed to be, 
\begin{equation}
    \frac{e^{g v_1}}{g} = \frac{e^{g v_2}}{g} + 2\Delta.
    \label{Eq:3.3.0.6}
\end{equation} 
For further calculations, let us take the magnitude of the shift $\Delta$ to be equal to $\frac{\lambda}{2 g}$, where $\lambda$ is some positive parameter. Thus, from Eq.~(\ref{Eq:3.3.0.6}) the relation between the null coordinates $v_1$ and $v_2$ can be written as,
\begin{eqnarray}
    v_1 = \frac{ln \big(e^{gv_2} + \lambda\big)}{g}.
    \label{Eq:3.3.0.7}
\end{eqnarray}
From the above mapping, it is clear that the LHS and RHS of the Eq.~(\ref{Eq:3.3.0.5}) are functions of $v_2$. So, taking the Fourier transform on both sides results in, 
\begin{equation}
    \begin{split}
    	\int_{-\infty}^{\infty}&\frac{dv_2}{\sqrt{2\pi}}e^{i\Omega v_2}\int_{0}^{\infty} \frac{dk}{\sqrt{2\pi}} \bigg[\hat{d}_1(k)\frac{e^{-ikv_1}}{\sqrt{2 k}} +  \hat{d}_1^\dagger(k)\frac{e^{ikv_1}}{\sqrt{2 k}}\bigg] \\ &=\int_{-\infty}^{\infty}\frac{dv_2}{\sqrt{2\pi}}e^{i\Omega v_2}\int_{0}^{\infty} \frac{dp}{\sqrt{2\pi}} \bigg[\hat{d}_2(p)\frac{e^{-ipv_2}}{\sqrt{2 p}} + \hat{d}_2^\dagger(p)\frac{e^{ipv_2}}{\sqrt{2 p}}\bigg].
    	\label{Eq:3.3.0.8}
    \end{split}
\end{equation}
Simplifying the Eq.~(\ref{Eq:3.3.0.8}) further results in $\hat{d}_2(\Omega)$ being expressed in terms of $R_1$ creation and annihilation operators as,
\begin{eqnarray}
    \hat{d}_2(\Omega) &=& \int_{0}^{\infty}dk \bigg[\hat{d}_1(k)\sqrt{\frac{ \Omega}{k}}\int_{-\infty}^{\infty}\frac{dv_2}{2\pi}e^{i(\Omega v_2-kv_1)} \nonumber\\
    &+& \hat{d}_1^\dagger(k)\sqrt{\frac{ \Omega}{k}}\int_{-\infty}^{\infty}\frac{dv_2}{2\pi}e^{i(\Omega v_2+ku_1)}\bigg]. 
    \label{Eq:3.3.0.9}
\end{eqnarray}
Upon comparing the above with the general expression of the $R_2$ annihilation operator in terms of operators of $R_1$ [see, Eq.~(\ref{Eq:3.2.0.3}) and Eq.~(\ref{Eq:3.2.0.4})], the Bogoliubov coefficients $\alpha^*_{21v}(k,\Omega)$ and $\beta^*_{21v}(k,\Omega)$ can be written as,
\begin{eqnarray}
    \alpha^*_{21v}(k,\Omega) =  \sqrt{\frac{ \Omega}{k}}\int_{-\infty}^{\infty}\frac{dv_2}{2\pi}e^{i(\Omega v_2-kv_1)},
    \label{Eq:3.3.0.10} \\
    \beta^*_{21v}(k,\Omega) =  -\sqrt{\frac{ \Omega}{k}}\int_{-\infty}^{\infty}\frac{dv_2}{2\pi}e^{i(\Omega v_2+kv_1)}.
    \label{Eq:3.3.0.11}
\end{eqnarray}
Now using the mapping [see Eq.~(\ref{Eq:3.3.0.7})], the Eq.~(\ref{Eq:3.3.0.10}) and the Eq.~(\ref{Eq:3.3.0.11}) can be evaluated to get non-zero Bogoliubov coefficients. Since $\beta^*_{21v}(k,\Omega) \neq 0$, which will be shown in the next subsection, we can confirm that the left-moving positive/negative frequency modes in $R_2$ can be expanded in terms of the left-moving positive and negative frequency modes of $R_1$ \cite{Padmanabhan_2010}. Also, since the positive frequency modes ($j_2 (q) \propto e^{-i \omega_q t_2}$) are proportional to the combination of the positive and negative frequency modes of $R_1$ due to the non-zero $\beta_{21v}(k,\Omega)$, the $R_1$ vacuum won't be a vacuum for $R_2$. In the subsequent subsections, first, we evaluate $\beta^*_{21v}(k,\Omega)$, then we proceed to compute the number density by multiplying ($\beta^*_{21v}(k,\Omega)\;\beta_{21v}(k,\Omega)$). Thus, we will explicitly show that for large frequencies $\Omega$, the computed number density distribution matches the Planckian distribution.

\subsubsection{Bogoliubov Coefficients:}
As discussed, showing $\beta^*_{21v}(k,\Omega) \neq 0$ is enough to validate the existence of a non-trivial Bogoliubov transformation between modes. So, in this subsection, we will evaluate the $\beta^*_{21v}(k,\Omega)$, which is represented by the integral Eq.~(\ref{Eq:3.3.0.11}) and show that it is non-zero. Since the integral [see Eq.~(\ref{Eq:3.3.0.11})] is evaluated with respect to $v_2$, we rewrite $v_1$ in terms of $v_2$ by using the relationship given by Eq.~(\ref{Eq:3.3.0.7}) to get,
\begin{equation}
    \beta^*_{21v}(k,\Omega) =  -\sqrt{\frac{ \Omega}{k}}\int_{-\infty}^{\infty}\frac{dv_2}{2\pi}e^{i(\Omega v_2+k\frac{\log{(e^{g v_2}+\lambda)}}{g})}.
    \label{Eq:3.3.1.1}
\end{equation}
Considering the acceleration parameter $g$ and shift parameter $\lambda$ as constants, and then substituting $e^{g v_2} =h$, such that $dv_2 = \frac{dh}{g h}$, the integral part, say ``$I(\Omega)$", in the above equation can be rewritten in terms of $h$ as,
\begin{equation}
    I(\Omega) =  \int_{0}^{\infty}\frac{dh}{g} \frac{h^{\frac{I \Omega}{h}} \big(h+\lambda\big)^{\frac{I k}{h}} }{h},
    \label{Eq:3.3.1.2}
\end{equation}
Note that the new limits of integration are from $0$ to $\infty$. Thus, evaluating the above integral $I(\Omega)$ and then combining it with the other constants, we get,
\begin{equation}
    \beta^*_{21v}(k,\Omega) =  -\frac{1}{2\pi g}\sqrt{\frac{ \Omega}{k}}\frac{ \lambda ^{\frac{i (k+\Omega )}{g}} \Gamma \left(\frac{i \Omega }{g}\right) \Gamma \left(-\frac{i (k+\Omega )}{g}\right)}{\Gamma \left(-\frac{i k}{g}\right)}. 
    \label{Eq:3.3.1.3}
\end{equation}
Also, the corresponding complex conjugate is,
\begin{equation}
    \beta_{21v}(k,\Omega) =  -\frac{1}{2\pi g}\sqrt{\frac{ \Omega}{k}}\frac{ \lambda ^{\frac{-i (k+\Omega )}{g}} \Gamma \left(\frac{-i \Omega }{g}\right) \Gamma \left(\frac{i (k+\Omega )}{g}\right)}{\Gamma \left(\frac{i k}{g}\right)}. 
    \label{Eq:3.3.1.4}
\end{equation}
Using these Bogoliubov coefficients, we can calculate the number density distribution for the vacuum state defined by the left-moving sector of the scalar field using $R_1$ modes in the $R_2$ frame. From the above expression for Bogoliubov coefficients, it can be concluded that the number density doesn't depend on the shift parameter $\lambda$.

\subsubsection{Particle Spectrum:}
To evaluate how the left-moving modes of the $R_1$ vacuum are perceived in the $R_2$ frame. We have to compute the number density for the $R_1$ vacuum. For this, we consider the expression for the expectation value of the number operator for the left-moving modes, and from it, we can, in turn, calculate the number density. The expectation value of the number operator $\hat{N}(k) \equiv \hat{d}_2^\dagger(k)\: \hat{d}_2(k)$ for any state with multiples modes (labeled with $k$) are \cite{birrell_davies_1982},
\begin{equation}
    \langle\hat{N}(k)\rangle = \int dk \; \hat{d}_2^\dagger(k) \:\hat{d}_2(k).
    \label{Eq:3.3.2.1}
\end{equation}
Thus, the expectation value of the $R_2$ number operator with respect to the $R_1$ vacuum state can be written as,
\begin{equation}
    \langle 0_{R_1} | \hat{N}(\Omega) | 0_{R_1}\rangle =  \int d\Omega \; \langle0_{R_1}| \hat{d}_2^\dagger(\Omega) \:\hat{d}_2(\Omega) | 0_{R_1}\rangle .
    \label{Eq:3.3.2.2}
\end{equation}
The above expression can be evaluated by substituting $R_2$ operators in terms of the $R_1$ operators and applying the commutation relations satisfied by the operators, $[\hat{d}_1(k),\:\hat{d}_1^\dagger(p)] = \delta(k-p)$ to get,
\begin{equation}
    \langle 0_{R_1}|\hat{N}(\Omega)|0_{R_1}\rangle =\int \frac{d\Omega}{2 \pi} \;\int_{0}^{\infty} dk \:|\beta_{21v}(\Omega,k)|^2
    \label{Eq:3.3.2.3}
\end{equation}
Now, from Eq.~(\ref{Eq:3.3.1.3}) and its complex conjugate Eq.~(\ref{Eq:3.3.1.4}), we can compute,
\begin{equation}
    |\beta_{21v}(k,\Omega)|^2 = \bigg[\frac{1}{e^{\frac{2 \pi  \Omega }{g}}-1}\bigg] \frac{ \sinh \left(\frac{\pi  k}{g}\right) \text{csch}\left(\frac{\pi  (k+\Omega )}{g}\right) }{2 \pi  g (k+\Omega ) e^{\frac{-\pi \Omega }{g}}}.
    \label{Eq:3.3.2.4}
\end{equation}
Thus the integral in the number density calculation can be written as,
\begin{equation}
    I(\Omega) = \int_{0}^{\infty} dk\; \frac{ \sinh \left(\frac{\pi  k}{g}\right) \text{csch}\left(\frac{\pi  (k+\Omega )}{g}\right) }{2 \pi  g (k+\Omega ) e^{\frac{-\pi \Omega }{g}}},
    \label{Eq:3.3.2.5}
\end{equation}
where the integrand can be represented by $F(k,\Omega)$. We note that the integrand tends to zero in the limit of $k \to 0$. The integrand $F(k,\Omega)$ is proportional to $k$. The integral, therefore, has no infrared divergence. In the limit $k \to \infty$, the integrand tends to zero again, and it is proportional to $\frac{1}{k}$. We, therefore, expect the integral to have logarithmic divergence. We investigate this below.
\begin{figure}[ht]
	\centering
	\includegraphics[scale=0.75]{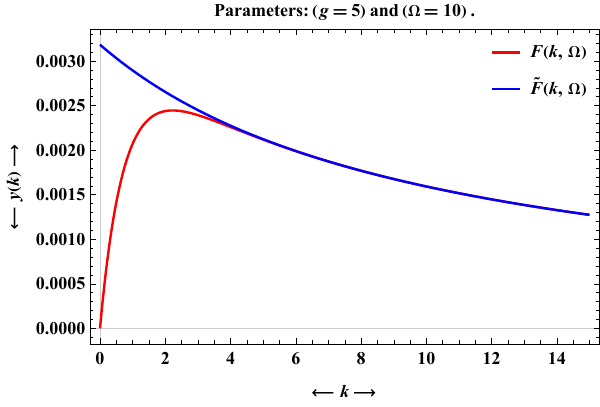}
	\caption{$F(k,\Omega)$ and  $\tilde{F}(k,\Omega)$,.\label{Fig:3}}
\end{figure}
Now, to solve this integral, we take the approximated form for the $F(k,\Omega)$, such that the approximated function $\tilde{F}(k,\Omega)$, coincides with $F(k,\Omega)$ for large $k$ since the main contribution of the integral comes from large $k$ (see Fig.~\ref{Fig:3}). Considering $F(k,\Omega)$, and using the fact that $\sinh{(x)} \approx \frac{e^{x}}{2}$, for large values of $x$, we can approximate $F(k,\Omega)$ as, 
\begin{equation}
    \begin{split}
        F(k,\Omega) \approx \tilde{F}(k,\Omega) = \frac{\frac{e^{\frac{\pi k}{g}}}{2} e^{\frac{\pi \Omega }{g}}}{2 \pi  g (k+\Omega ) \frac{e^{\frac{\pi (k+\Omega)}{g}}}{2}}.
    \end{split}
    \label{Eq:3.3.2.6}
\end{equation}
On simplification, we get,
\begin{equation}
    \tilde{F} (k,\Omega) = \frac{1}{2 \pi g (k+\Omega )}.
    \label{Eq:3.3.2.7}
\end{equation}
Clearly, from the Fig.~\ref{Fig:3}, we can see $\tilde{F}(k,\Omega)$ bounds $F(k,\Omega)$ for large values of $k$ and matches perfectly with $F(k,\Omega)$ for large values of $k$.
Thus, we can write the new integral as,
\begin{equation}
     \tilde{I}(\Omega) = \int_{0}^{\infty} dk \: \tilde{F}(k,\Omega).
    \label{Eq:3.3.2.8}
\end{equation}
Therefore, the number density can be computed as,
\begin{equation}
    \langle 0_{R_1}|\hat{N}(\Omega)|0_{R_1}\rangle = \int \frac{d\Omega}{2 \pi} \bigg[\frac{1}{e^{\frac{2\pi \Omega}{g}} - 1}\bigg]\int_{0}^{\infty} dk \frac{1}{2\pi g }\frac{1}{(k+\Omega)}.
    \label{Eq:3.3.2.9}
\end{equation}
The integral in the above equation is logarithmically divergent, and this divergent part has units of spatial volume. We assume that this infinity is associated with the infinite spatial volume, and these are well-known infinities that arise even when one calculates the particle density in Rindler spacetime arising from the Minkowski vacuum. Since the number density $\langle \hat{n}(\Omega) \rangle$, is given as,
\begin{equation}
    \langle 0_{R_1}|\hat{N}(\Omega)|0_{R_1}\rangle = \frac{1}{V}\int \frac{d\Omega}{2 \pi}\:\langle 0_{R_1}| \hat{n}(\Omega)|0_{R_1}\rangle.
    \label{Eq:3.3.2.10}
\end{equation} 
Thus from Eq.~(\ref{Eq:3.3.2.10}) and Eq.~(\ref{Eq:3.3.2.8}), we get the number density as,
\begin{equation}
    \langle \hat{n}(\Omega)\rangle = \frac{1}{\big[e^{\frac{2\pi\Omega}{g}} - 1\big]},
    \label{Eq:3.3.2.11}
\end{equation}
which is a Planckian distribution, as expected for the Scalar fields. 
Thus, it is very clear from the above analysis that modes are selectively excited in special scenarios, such as in our case, where we have considered null-shifted wedges. In particular, only the left-moving modes are excited in the case under consideration.  This is a case of informative thermalization since the particle content reveals the causal positioning of the superset ($R_1$) that yields the given selective thermal state.

\section{ Chiral Excitation  for massless Dirac fields\label{Sec-4}}
We now explore the excitations of fermionic fields since these are more intimately connected to spacetime points than massless scalar fields (due to the additional structures called spin connections). Since the Rindler spacetime works very well as a toy model for the black hole scenarios involving horizons, we set up Dirac fields in Rindler spacetime to capture the effect of null-shifted horizons on Dirac fields.  We consider Minkowski and Rindler to be the background spacetimes. Note that we continue the notation where the Minkowski coordinates are labeled using $(T, X)$ and the Rindler using $(t,x)$ and specifically for the $R_1$ frame, it is $(t_1,x_1)$, and for $R_2$ it is $(t_2,x_2)$.
\subsection{Dirac Fields in Minkowski spacetime \label{Subsec-4.1}}
The Dirac equation for massless spin $1/2$ particle in $(1+1)$ dimensional Minkowski spacetime can be written as, 
\begin{equation}
	i \gamma^\mu \partial_\mu \psi = 0.
	\label{Eq:4.1.0.1}
\end{equation}
Here, we can choose the gamma matrices in chiral representation for two-dimensional Minkowski spacetime as \cite{RobertMann_1991, Flouris.98.155419},
\begin{equation}
	\begin{matrix}
		\gamma^0 = 
		\begin{bmatrix}
			0 & 1 \\
			1 & 0
		\end{bmatrix}, & 
		\gamma^1 = 
		\begin{bmatrix}
			0 & -1 \\
			1 & 0
		\end{bmatrix}
	\end{matrix}.
	\label{Eq:4.1.0.2}
\end{equation}
So that the above gamma matrices obey the Clifford algebra,
\begin{equation}
	{\gamma^\mu \gamma^\nu + \gamma^\nu \gamma^\mu} = 2 \eta^{\mu\nu}.
	\label{Eq:4.1.0.3}
\end{equation}
Moreover, $(\gamma^0)^2 = I$ and $(\gamma^1)^2 = -I$ are additional constraints imposed due to the choice of chiral representation (see Appendix B.3 of Ref.~\cite{collas2019dirac}). Also, the chirality operator for (1+1) dimensions is defined as \cite{Palash.B.3549729},
\begin{equation}
    \gamma^5:=\gamma^0 \gamma^1.
    \label{Eq:4.1.0.4}
\end{equation}
Now, substituting the Eq.~(\ref{Eq:4.1.0.2}) in Eq.~(\ref{Eq:4.1.0.1}), we get the expanded form of the Dirac equation,
\begin{equation}
	\big[i \gamma^0 \partial_0 + i \gamma^1 \partial_1\big] \psi = 0.
	\label{Eq:4.1.0.5}
\end{equation}
This equation is satisfied by a wave function in the form of a two-by-two-column matrix,
\begin{equation}
    \psi = \begin{bmatrix}
        \psi_{(+)} \\
        \psi_{(-)}
    \end{bmatrix}.
    \label{Eq:4.1.0.6}
\end{equation}
Now, in the case of massless particles, Eq.~(\ref{Eq:4.1.0.6}) in Eq.~(\ref{Eq:4.1.0.5}) gives
two decoupled differential equations,
\begin{equation}
    \frac{\partial \: \psi_{(-)}}{\partial T} - \frac{\partial \: \psi_{(-)}}{\partial X} = 0,
    \label{Eq:4.1.0.7}
\end{equation}
and,
\begin{equation}
    \frac{\partial \: \psi_{(+)}}{\partial T} + \frac{\partial \: \psi_{(+)}}{\partial X} = 0,
    \label{Eq:4.1.0.8}
\end{equation}
which can be easily solved to get $\psi$.
We can, therefore write the mode expansion of the Dirac field $\Psi$ as, 
\begin{equation}
    \begin{split}
        \Psi = \frac{1}{\sqrt{2 \pi}} \int_{0}^{\infty} & dk\bigg(a(k)
	\begin{bmatrix}
		e^{i k(T-X)}  \\ 0
	\end{bmatrix}
 +c^\dagger(k)
 \begin{bmatrix}
		e^{-i k(T-X)} \\ 0
	\end{bmatrix}\\
 &+b(k)
	\begin{bmatrix}
		0  \\ e^{-i k(T+X)}
	\end{bmatrix}
 +d^\dagger(k)
 \begin{bmatrix}
		0 \\ e^{i k(T+X)}
	\end{bmatrix}
 \bigg)
    \end{split}
    \label{Eq:4.1.0.9}
\end{equation}

The corresponding plane wave solutions in lightcone coordinates  $\big[(U=T-X) \text{ and } (V = T+X)\big]$ are,
\begin{equation}
	\Psi = \psi_U + \psi_V,
	\label{Eq:4.1.0.10}
\end{equation}
where $\psi_U$ and $\psi_V$ are decoupled and correspond to the right-moving sector and the left-moving sector of the field, respectively. For the rest of the calculations in this article, we consider the simplest case of Dirac fields, i.e., the real representation of the Dirac field commonly known as the Majorana representation. In the case of Majorana fields, we have to impose the constraint that the field $\Psi$ has to be real, i.e.,
\begin{equation}
    \Psi = \Psi^*
    \label{Eq:4.1.0.11}
\end{equation}
where the ``$*$" denotes complex conjugate. It has to be noted that a Majorana field describes particles with properties similar to Dirac particles except that particles and anti-particles are identical \cite{Freedman_Van_2012}. Now, with these considerations, the mode expansion for the respective right-moving and left-moving sectors of a Majorana field can be written as,
\begin{equation}
	\psi_{U} =\frac{1}{\sqrt{2\pi}} \int_{0}^{\infty}dk
	\bigg(a(k)\begin{bmatrix}
		e^{-i k U}\\
		0
	\end{bmatrix}+
	a^\dagger(k)\begin{bmatrix}
		e^{i k U}\\
		0
	\end{bmatrix}\bigg),
	\label{Eq:4.1.0.12}
\end{equation}
and,
\begin{equation}
	\psi_{V} = \frac{1}{\sqrt{2\pi}} \int_{0}^{\infty}dk
	\bigg(b(k)\begin{bmatrix}
		0\\
		e^{-i k V}
	\end{bmatrix}+
	b^\dagger(k)\begin{bmatrix}
		0\\
		e^{i k V}
	\end{bmatrix}\bigg),
	\label{Eq:4.1.0.13}
\end{equation}
where $1/ \sqrt{2\pi}$ is the normalization factor determined from the inner product defined on the constant time hypersurface. It is important to note that $k$ is always positive, and the operators $\hat{a}_1^\dagger(k),\hat{b}_1^\dagger(k)$ and $\hat{a}_1(k),\hat{b}_1(k)$ are the usual creation and annihilation operators of the Dirac field (Majorana), and they obey the standard anti-commutation relations:
\begin{equation}
    \begin{split}
        \{\hat{a}_1(k),\hat{a}_1^\dagger(k')\} =& \delta(k-k'),\\
        \{\hat{a}_1(k),\hat{a}_1(k')\} =& 0,\\
        \{\hat{a}_1^\dagger(k),\hat{a}_1^\dagger(k')\} =& 0.\\
    \end{split}
    \label{Eq:4.1.0.14}
\end{equation}
Similar is the case with the operators $\hat{b}_1^\dagger(k)$ and $\hat{b}_1(k)$. In the next section, we discuss describing a Dirac field (Majorana) in the Rinlder spacetime.

\subsection{Dirac Fields in Rindler spacetime \label{Subsec-4.2}}
Dirac fields in the Rindler metric can be studied using the vierbein formalism \cite{Frankel_2011,Fecko_2006}. Thus, we can write,
\begin{equation}
	\eta_{AB} = e^\mu_A e^\nu_B g_{\mu\nu},
	\label{Eq:4.2.0.1}
\end{equation}
where the Zweibeins $e_A^\mu$ are used to project vectors from a coordinated frame (Global) to a non-coordinated frame (local Lorentz) \cite{RobertMann_1991}.
For the case of the Rindler metric, the Zweibeins can be written as,
\begin{equation}
	e^\mu_A = 
	\begin{bmatrix}
		e^{-g x} & 0\\
		0 & e^{-g x}
	\end{bmatrix},
	\label{Eq:4.2.0.2}
\end{equation}
and this determines the frame for the Dirac particle. Another important thing for writing down the Dirac equation in Rindler spacetime is the spacetime-dependent gamma matrix, $\tilde{\gamma}^\mu$. It can be computed from the  Zweibeins and the normal flat spacetime gamma matrix, and the connection is given below as,
\begin{equation}
	\tilde{\gamma}^\mu = e^\mu_A \gamma^A.
	\label{Eq:4.2.0.3}
\end{equation}
Also, we note that the inner product obeyed by the Dirac Spinors can be written as \cite{Collas2019_2},
\begin{equation}
	\langle \phi|\psi\rangle = \int_{\Sigma} \overline{\phi}\; \overline{\gamma}^\mu n_\mu\; \psi \:d\Sigma,
	\label{Eq:4.2.0.4}
\end{equation}
where $\overline{\phi} = \phi^\dagger \gamma^0$, $n$ is the future directed normal to the hypersurface $\Sigma$ and $d\Sigma$ is the invariant volume element on $\Sigma$. 
Following the analysis done in references \cite{RobertMann_1991,Collas2019_1}, the general form for a Dirac equation in an arbitrary $(1+1)$ dimensional spacetime can be written as,
\begin{equation}
	\bigg[i \gamma^A e^\mu_A \partial_\mu + \frac{1}{2}\gamma^A \frac{1}{\sqrt{-\overline{g}}}\partial_\mu\big(\sqrt{-\overline{g}} e^\mu_A\big)\bigg]\Psi = 0,
	\label{Eq:4.2.0.5}
\end{equation}
where $\overline{g}$ is the determinant of the metric. Substituting the parameters of the Rindler metric on Eq.~(\ref{Eq:4.2.0.5}) results in,
\begin{equation}
	\bigg[i e^{-g x}\gamma^0 \partial_0 + i e^{-g x} \gamma^1 \partial_1 + \frac{1}{2} i g e^{-g x} \gamma^1\partial_1\bigg]\Psi = 0.
	\label{Eq:4.2.0.6}
\end{equation}
From Eq.~(\ref{Eq:2.1.1.9}), we note that the Rindler metric does not depend on $t$. Following the calculations in \cite{,Collas2019_3,RobertMann_1991}, the trial solution for positive frequency mode $\Psi_+ (k)$  are,
\begin{equation}
	\Psi_{+}(k) = e^{-i \omega_k t}
	\begin{bmatrix}
		f_1(x) \\ 
		f_2(x)
	\end{bmatrix},
	\label{Eq:4.2.0.7}
\end{equation} 
where $f_1(x) \text{ and } f_2(x)$ are the space dependent part of the solution. 
Substituting this in Eq.~(\ref{Eq:4.2.0.6}), we obtain, 
\begin{equation}
	2 i \frac{d\:f_2(x)}{dx}- \big(2 \omega_k - i g\big)f_2(x) = 0,
	\label{Eq:4.2.0.8}
\end{equation}
and
\begin{equation}
	2 i \frac{d\:f_1(x)}{dx} + \big(2 \omega_k + i g\big)f_1(x) = 0.
	\label{Eq:4.2.0.9}
\end{equation}
The above differential equations are solved to get the spatial part. It is then combined with the temporal part to get the total solution and is normalized using the inner product defined by Eq.~(\ref{Eq:4.2.0.4}). Since $f_1(x)$ and $f_2(x)$ are decoupled, we can write the complete positive mode solution as,
\begin{equation}
	\Psi_+(k) = \frac{e^{\frac{-g x}{2}}}{\sqrt{2 \pi}}\bigg(c(k)
	\begin{bmatrix}
		e^{-i k(t-x)} \\0
	\end{bmatrix}
 + d(k)
 \begin{bmatrix}
		0  \\ e^{-i k(t+x)}
	\end{bmatrix}
 \bigg),
	\label{Eq:4.2.0.10}
\end{equation}
where we have taken the on-shell condition of $\omega_k=|k|$ for massless fields and also $1/\sqrt{2 \pi}$ as the normalization constant. A similar analysis provides the mode expansion for the negative mode solutions as,
\begin{equation}
	\Psi_-(k) = \frac{e^{\frac{-g x}{2}}}{\sqrt{2 \pi}} \bigg(c^\dagger(k)
	\begin{bmatrix}
		e^{i k(t-x)}  \\ 0
	\end{bmatrix}
 +d^\dagger(k)
 \begin{bmatrix}
		0 \\ e^{i k(t+x)}
	\end{bmatrix}
 \bigg).
	\label{Eq:4.2.0.11}
\end{equation}
Combining the two and expressing them as right-moving and left-moving solutions as in Eq.~(\ref{Eq:4.1.0.4}), we get,
\begin{equation}
	\begin{split}
		\Psi =& \frac{e^{\frac{-g x}{2}}}{\sqrt{2 \pi}}\int_{0} ^{\infty}dk
		\bigg(c(k)
		\begin{bmatrix}
			e^{-i k(t-x)} \\
			0
		\end{bmatrix} + c^\dagger (k)
		\begin{bmatrix}
			e^{i k(t-x)} \\
			0
		\end{bmatrix}\bigg)\\ 
		&+ \frac{e^{\frac{-g x}{2}}}{\sqrt{2 \pi}}\int_{0} ^{\infty}dk\bigg(d(k)
		\begin{bmatrix}
			0 \\
			e^{-i k(t+x)}
		\end{bmatrix} + d^\dagger (k)
		\begin{bmatrix}
			0 \\
			e^{i k(t+x)}
		\end{bmatrix}
		\bigg).
	\end{split}
	\label{Eq:4.2.0.12}
\end{equation}
Note: The above mode expansion is for a Dirac field operator in a general Rindler spacetime setting; till now, we haven't specified any particular reference frame. Below, when used for calculations in a specific Rindler frame, the $x$ and $t$ will be replaced with the frame's coordinates $(t_i, x_i)$, where the label ``$i$" indicates the corresponding frame. 

\subsection{Thermal spectrum of fermions in Rindler spacetime from Minkowski vacuum state\label{Subsec-4.3}}
To check for consistency, we will briefly derive the well-known result of the thermal spectrum for fermions in $R_1$ spacetime from the Minkowski Vacuum for the Dirac fields in Rindler spacetime.  The Eq.~(\ref{Eq:4.1.0.5}) gives the right-moving modes of the Dirac field in terms of Minkowski modes, and similarly, from Eq.~(\ref{Eq:4.2.0.12}), we get the right-moving modes in terms of $R_1$. As already discussed earlier, we take the ansatz that the modes of $R_1$ are related to the modes of Minkowski through a Bogoliubov transformation of the following form:
\begin{equation}
     h_1(q) = \int_{0}^{\infty} dk \bigg(\alpha_{10U}(q,k)f_M(k) + \beta_{10U}(q,k)f_M^*(k)\bigg),  \label{Eq:4.3.0.1}
\end{equation}
\begin{equation}
	f_M(k) = \int_{0}^{\infty} dq \bigg(\alpha^*_{10U}(q,k)h_1(q) + \beta_{10U}(q,k)h_1^*(q)\bigg),
	\label{Eq:4.3.0.2}
\end{equation}
where $f_M$ are the Minkowski modes and $h_1$ are the $R_1$ modes respectively. Note that the subscript ``$U$" in the Bogoliubov coefficients [Eg, $\alpha_{10U}(q,k)$] implies that it is for the right-moving modes, and the subscript ``$10$" implies the Bogoliubov coefficient is for the modes of $R_1$ and Minkowski.
Based on the above ansatz [see Eq.~(\ref{Eq:4.3.0.1}) and Eq.~(\ref{Eq:4.3.0.2})] the operators of $R_1$ can be expressed in terms of Minkowski as,
\begin{equation}
      \hat{c}_1(p) = \int_{0}^{\infty} dk \bigg(\alpha^*_{10U}(p,k)\hat{a}_M(k) + \beta^*_{10U}(p,k)\hat{a}_M^\dagger(k)\bigg), \label{Eq:4.3.0.3}
\end{equation}
\begin{equation}
       \hat{c}_1^\dagger(p) = \int_{0}^{\infty} dk \bigg(\alpha_{10U}(p,k)\hat{a}_M^\dagger(k) + \beta_{10U}(p,k)\hat{a}_M(k)\bigg).
	\label{Eq:4.3.0.4}
\end{equation}
Now, coming back to the field operators in $R_1$ and Minkowski, it is clear that at any point in the right-wedge of the $R_1$, the field operator of the right moving sector can be independently expressed using the right modes of $R_1$ or right modes of Minkowski as a basis \cite{Valdivia-Mera.200109869V}, i.e.,
\begin{equation}
		\begin{split}
			&\Psi(x_\mu)=\\& \int_0^\infty  d k\frac{e^{\frac{-g x_1}{2}}}{\sqrt{2\pi}}\bigg(c_{1}(k)\begin{bmatrix}
				e^{-i k(t_1-x_1)}\\0
			\end{bmatrix}+c_{1}^\dagger (k)\begin{bmatrix}
				e^{i k(t_1-x_1)}\\0
			\end{bmatrix}\bigg)\\
			&=\int_0^\infty d p \frac{1}{\sqrt{2\pi}}\bigg(a_{M}(p)\begin{bmatrix}
				e^{-i p(T-X)}\\0
			\end{bmatrix}+a_{M}^\dagger(p)\begin{bmatrix}
				e^{i p(T-X)}\\0
			\end{bmatrix}\bigg).
		\end{split}
            \label{Eq:4.3.0.5}
	\end{equation}
For $t_1=T=0$, Cauchy hypersurface, the above equation becomes,
\begin{equation}
		\begin{split}
			&\int_0^\infty  d k\frac{e^{\frac{-g x_1}{2}}}{\sqrt{2\pi}}\bigg(c_{1}(k)\begin{bmatrix}
				e^{i k x_1}\\0
			\end{bmatrix}+c_{1}^\dagger (k)\begin{bmatrix}
				e^{-i k x_1}\\0
			\end{bmatrix}\bigg)\\
			&=\int_0^\infty d p \frac{1}{\sqrt{2\pi}}\bigg(a_{M}(p)\begin{bmatrix}
				e^{i p X}\\0
			\end{bmatrix}+a_{M}^\dagger(p)\begin{bmatrix}
				e^{-i p X}\\0
			\end{bmatrix}\bigg).
		\end{split}
            \label{Eq:4.3.0.6}
	\end{equation}
It is clear from Eq.~(\ref{Eq:2.1.1.2}), $X = \frac{e^{g x_1}}{g}$, for $t_1=0$, thus the LHS and RHS of Eq.~(\ref{Eq:4.3.0.6}) are functions of $x_1$. Taking the Fourier transform on both sides, we get,
\begin{equation}
		\begin{split}
			\int_0^\infty  d k&\int_{-\infty}^\infty\frac{dx_1}{2\pi}\bigg(c_{1}(k)
				e^{i (k+\Omega) x_1} + c_{1}^\dagger (k) e^{i(\Omega- k) x_1}\bigg)\\
			&=\int_0^\infty d p \bigg[ \bigg(\int_{-\infty}^\infty\frac{dx_1}{2\pi} e^{\frac{g x_1}{2}}  e^{i\Omega x_1} e^{i p \frac{e^{g x_1}}{g}}\bigg) a_{M}(p)\\
   &+ \bigg(\int_{-\infty}^\infty\frac{dx_1}{2\pi} e^{\frac{g x_1}{2}} e^{i\Omega x_1}  e^{-i p \frac{e^{g x_1}}{g}}\bigg) a_{M}^\dagger(p)\bigg].
		\end{split}
            \label{Eq:4.3.0.7}
	\end{equation}
For $\Omega>0$, i.e., positive, we can evaluate the LHS to be $c_{1}^\dagger (\Omega)$ and then compare the Eq.~(\ref{Eq:4.3.0.7}) with Eq.~(\ref{Eq:4.3.0.4}), to get,
\begin{equation}
    \begin{split}
        \alpha_{10U}(\Omega,p) = \int_{-\infty}^\infty\frac{dx_1}{2\pi} e^{\frac{g x_1}{2}} e^{i\Omega x_1}  e^{-i p \frac{e^{g x_1}}{g}},
    \end{split}
    \label{Eq:4.3.0.8}
\end{equation}
and,
\begin{equation}
    \begin{split}
        \beta_{10U}(\Omega,p) = \int_{-\infty}^\infty\frac{dx_1}{2\pi} e^{\frac{g x_1}{2}}  e^{i\Omega x_1} e^{i p \frac{e^{g x_1}}{g}}.
    \end{split}
    \label{Eq:4.3.0.9}
\end{equation}
Since we are interested in the number density, we will evaluate $\beta_{10U}(\Omega,p)$. To evaluate $\beta_{10U}(\Omega,p)$ from the Eq.~(\ref{Eq:4.3.0.9}), we will substitute $e^{g x_1} = h$, such that $dx_1 = \frac{dh}{gh}$. Thus rewriting everything in terms of $h$ and applying new limits to the integral, we get,
\begin{equation}
    \beta_{10U}(\Omega,p)=\frac{\left(-\frac{i p}{g}\right)^{-\frac{1}{2}-\frac{i \Omega }{g}} \Gamma \left(\frac{i \Omega }{g}+\frac{1}{2}\right)}{2 \pi  g}.
    \label{Eq:4.3.0.10}
\end{equation}
Now, in order to find the number density, i.e., \mbox{$N = \int |\beta|^2$}, we will modify the form of the above expression Eq.~(\ref{Eq:4.3.0.10}), such that the resulting expression, when $\beta_{10U}(\Omega,p)$ and $\beta^*_{10U}(\Omega,p)$ are multiplied, can be simplified easily. Thus the modified form of the Bogoliubov coefficient $\beta_{10U}(\Omega,p)$ is,
\begin{equation}
    \beta_{10U}(\Omega,p) =  \frac{e^{\frac{i \pi}{4}}e^{-\frac{\pi \Omega}{2 g}}\left(\frac{p}{g}\right)^{-\frac{1}{2}-\frac{i \Omega }{g}} \Gamma \left(\frac{1}{2}+\frac{i \Omega }{g}\right)}{2 \pi  g}.
    \label{Eq:4.3.0.11}
\end{equation}
From the Eq.~(\ref{Eq:4.3.0.11}), the expectation of particle number can be found as,
\begin{equation*}
    \begin{split}
        \langle& N(\Omega) \rangle = \int_{0}^{\infty} \frac{d\Omega}{2\pi} \int_{0}^{\infty}dp\;|\beta_{10U}(\Omega,p)|^2,\\
        &= \int_{0}^{\infty} \frac{d\Omega}{2\pi} \int_{0}^{\infty}dp \bigg(\frac{ e^{-\frac{\pi \Omega}{ g}}  \left(\frac{p}{g}\right)^{-1}  |\Gamma \left(\frac{1}{2}+\frac{i \Omega }{g}\right)|^2 }{4 \pi^2  g^2} \bigg).
    \end{split} 
\end{equation*}
\begin{equation*}
    \langle N(\Omega) \rangle = \int_{0}^{\infty} \frac{d\Omega}{2\pi}\frac{1}{\big[ e^{\frac{2 \pi  \Omega }{g}}+1 \big]} \underbrace{\int_{0}^{\infty} d p \:\frac{1}{p}\:\frac{1}{2\pi g }}_{\delta(0)\rightarrow \: \text{divergent factor}},
\end{equation*}
Thus, the number density is,
\begin{equation*}
    \langle n(\Omega) \rangle = \frac{1}{\big[ e^{\frac{2 \pi  \Omega }{g}}+1 \big]}.
    \label{Eq:4.3.0.12}
\end{equation*}
The above result affirms the Unruh effect for the case of Dirac Fields \cite{R.auregui.43.3979, Soffel1935}. The subsequent sections provide a detailed discussion of the primary focus of this work, i.e., the selective excitation of the Dirac fields. 

\subsection{Negligible Excitations of Right handed modes:\label{Subsec-4.4}}
In this and the following subsections, we start with the Dirac field in vacuum for $R_1$ and evaluate the particle spectrum or excitations for $R_2$. For the study of excitations of right-moving modes, the same line of analysis is adopted as in Sec.~\ref{Sec-3}, where the case of a free massless scalar field was considered. Now, to study any selective thermalization phenomenon in the case of Dirac particles, we again consider the case where $R_2 \subset R_1$ (see Fig.~\ref{Fig:1}). In the below section, we will answer how $R_2$ will perceive a Dirac (Majorana) field in a vacuum state defined by $R_1$, and from it, we will understand if any selective excitation exists for the case of Dirac particles. Now, once again, we take the ansatz that the modes of $R_2$ represented by $h_2$ are related to the modes of $R_1$ represented by $h_1$, through a Bogoliubov transformation of the following form,
\begin{equation}
     h_2(q) = \int_{0}^{\infty} dk \bigg(\alpha_{21u}(q,k)h_1(k) + \beta_{21u}(q,k)h_1^*(k)\bigg),  \label{Eq:4.4.0.1}
\end{equation}
\begin{equation}
	h_1(k) = \int_{0}^{\infty} dq \bigg(\alpha^*_{21u}(q,k)h_2(q) + \beta_{21u}(q,k)h_2^*(q)\bigg).
	\label{Eq:4.4.0.2}
\end{equation}
Note that the subscript ``$u$" in the Bogoliubov coefficients [Eg, $\alpha_{21u}(q,k)$] implies that it is for the right moving modes (for left moving modes, the subscript ``$u$" will be replaced by ``$v$"). Also, the subscript ``$21$" implies the Bogoliubov coefficient is for the modes of $R_2$ and $R_1$. Based on the above ansatz, the creation/annihilation operators of $R_2$ can be expressed in terms of $R_1$ as,
\begin{equation}
      \hat{c}_2(p) = \int_{0}^{\infty} dk \bigg(\alpha^*_{21u}(p,k)\hat{c}_1(k) + \beta^*_{21u}(p,k)\hat{c}_1^\dagger(k)\bigg), 
      \label{Eq:4.4.0.3}
\end{equation}
\begin{equation}
       \hat{c}_2^\dagger(p) = \int_{0}^{\infty} dk \bigg(\alpha_{21u}(p,k)\hat{c}_1^\dagger(k) + \beta_{21u}(p,k)\hat{c}_1(k)\bigg).
	\label{Eq:4.4.0.4}
\end{equation}
Since the probe field is massless, the on-shell conditions, i.e., $|\omega_k| = k$, are used to express the mode expansion of the field operator. In the common patch of $R_1$ and $R_2$, the field operator of the right moving sector is expanded using the right modes as,
\begin{equation}
		\begin{split}
			&\int_0^\infty d k\frac{e^{\frac{-g x_1}{2}}}{\sqrt{2\pi}}\bigg(c_{1}(k)\begin{bmatrix}
				e^{-ik(t_1-x_1)}\\0
			\end{bmatrix}+c_{1}^\dagger(k)\begin{bmatrix}
				e^{ik(t_1-x_1)}\\0
			\end{bmatrix}\bigg)\\
			&=\int_0^\infty d p \frac{e^{\frac{-g x_2}{2}}}{\sqrt{2\pi}}\bigg(c_{2}(p)\begin{bmatrix}
				e^{-ip(t_2-x_2)}\\0
			\end{bmatrix}+c_{2}^\dagger(p)\begin{bmatrix}
				e^{i p(t_2-x_2)}\\0
			\end{bmatrix}\bigg).
		\end{split}
        \label{Eq:4.4.0.5}
	\end{equation}
 In terms of null coordinates, the above becomes, 
 \begin{equation}
		\begin{split}
			&\int_0^\infty dk\frac{e^{\frac{-g }{4}(v_1-u_1)}}{\sqrt{2\pi}}\bigg(c_{1}(k)\begin{bmatrix}
				e^{-i k u_1}\\0
			\end{bmatrix}+c_{1}^\dagger(k)\begin{bmatrix}
				e^{i k u_1}\\0
			\end{bmatrix}\bigg)\\
			&=\int_0^\infty dp \frac{e^{\frac{-g }{4}(v_2-u_2)}}{\sqrt{2\pi}}\bigg(c_{2}(p)\begin{bmatrix}
				e^{-i p u_2}\\0
			\end{bmatrix}+c_{2}^\dagger(p)\begin{bmatrix}
				e^{i p u_2}\\0
			\end{bmatrix}\bigg).
		\end{split}
        \label{Eq:4.4.0.6}
	\end{equation}
To get the expressions for the Bogoliubov coefficients, we take the Fourier transform on both sides, i.e., we multiply both LHS and RHS of Eq.~(\ref{Eq:4.4.0.6}) with
\begin{equation}
    \int_{-\infty}^{\infty} \frac{d u_2}{2 \pi} e^{\frac{g }{4}(v_2-u_2)}
    \begin{bmatrix}
		e^{i \Omega u_2} & 0
	\end{bmatrix}.
    \label{Eq:4.4.0.7}
\end{equation} 
Such that the RHS of Eq.~(\ref{Eq:4.4.0.6}) can be written as,
 \begin{equation}
     \begin{split}
         \int\limits_0^\infty dp \bigg(c_{2}(p)s\int_{-\infty}^{\infty}&\frac{du_2}{2\pi}e^{i(\Omega - p)u_2}\\&+c_{2}^\dagger(p)\int_{-\infty}^{\infty}\frac{du_2}{2\pi}e^{i(\Omega + p)u_2}\bigg),
     \end{split}
     \label{Eq:4.4.0.8}
 \end{equation}
  and similarly, the LHS as,
 \begin{equation}
		\begin{split}
			\int_0^\infty dk&\bigg(c_{1}(k)\int_{-\infty}^{\infty}\frac{du_2}{2\pi}e^{\frac{g}{4}(v_2-u_2-v_1+u_1)}e^{i(\Omega u_2 - k u_1)}\\
            &+c_{1}^\dagger(k)\int_{-\infty}^{\infty}\frac{du_2}{2\pi}e^{\frac{g}{4}(v_2-u_2-v_1+u_1)}e^{i(\Omega u_2 + k u_1)}\bigg).
		\end{split}
        \label{Eq:4.4.0.9}
	\end{equation}
 The RHS can be simplified to get,
 \begin{equation}
     \int_0^\infty dp\bigg(c_{2}(p)\delta(\Omega-p)+c_{2}^\dagger(p)\delta(\Omega+p)\bigg)=c_{2}(\Omega),
     \label{Eq:4.4.0.10}
 \end{equation}
 for $p>0$ and $\Omega>0$. From the Bogoliubov transformations [see Eq~(\ref{Eq:4.4.0.1}) and Eq~(\ref{Eq:4.4.0.2})], the operators of $R_1$ and $R_2$ are related by Eq~(\ref{Eq:4.4.0.3}) and Eq~(\ref{Eq:4.4.0.4}), thus from these relations and from Eq~(\ref{Eq:4.4.0.10}) and Eq~(\ref{Eq:4.4.0.9}), the Bogoliubov coefficients can be written as,
 \begin{equation}
     \alpha_{21u}^*(\Omega,k)=\int_{-\infty}^\infty\frac{du_2}{2\pi}e^{\frac{g}{4}(v_2-u_2-v_1+u_1)}e^{i(\Omega u_2 - k u_1)},
     \label{Eq:4.4.0.11}
 \end{equation}
 and 
 \begin{equation}
     \beta_{21u}^*(\Omega,k)=\int_{-\infty}^\infty\frac{du_2}{2\pi}e^{\frac{g}{4}(v_2-u_2-v_1+u_1)}e^{i(\Omega u_2 + k u_1)}.
     \label{Eq:4.4.0.12}
 \end{equation}
To evaluate $\beta_{21u}^*(\Omega,k)$, consider a constant time hypersurface $t_2=0$ on the common patch of $R_1$ and $R_2$. For the case described by the Fig.~\ref{Fig:1}, from the Eq.~(\ref{Eq:3.2.0.6}) we get $u_1 = u_2$ and from Eq.~(\ref{Eq:2.1.1.8.a}) and Eq.~(\ref{Eq:2.1.1.8.b}), at $t_2=0$ we get $v_2 = -u_2$. Now, using the above facts and the Eq.~(\ref{Eq:3.3.0.7}), we can rewrite the integral in Eq.~(\ref{Eq:4.4.0.12}) as,
\begin{equation}
     \beta_{21u}^*(\Omega,k)=\int_{-\infty}^\infty\frac{du_2}{2\pi}e^{\frac{-g}{4}(u_2+ln{[e^{-gu_2}+\lambda]})}e^{i(\Omega + k )u_2 }.
     \label{Eq:4.4.0.13}
 \end{equation}
Substituting $h= e^{- gu_2}$, such that $du_2 = -\frac{dh}{gh}$ and applying the new limits for integration we get,
\begin{equation}
     \beta_{21u}^*(\Omega,k)= \frac{1}{2\pi g}\int_{0}^\infty du_2 \bigg[ \frac{h}{h+\lambda} \bigg]^{\frac{1}{4}} h^{\frac{-i (\Omega + k)}{g} - 1} .
     \label{Eq:4.4.0.14}
 \end{equation}
Solving this, we get,
\begin{equation}
     \beta_{21u}^*(\Omega,k)= \frac{\lambda^{\frac{-i(k+\Omega)}{g}}}{2\pi g} \frac{\Gamma \left(\frac{i (k +\Omega )}{g}\right) \Gamma \left(\frac{1}{4}-\frac{i (k +\Omega )}{g}\right)}{\Gamma \left(\frac{1}{4}\right)} .
     \label{Eq:4.4.0.15}
 \end{equation}
Also, its complex conjugate is,
\begin{equation}
     \beta_{21u}(\Omega,k)= \frac{\lambda^{\frac{i(k+\Omega)}{g}}}{2\pi g} \frac{\Gamma \left(\frac{-i (k +\Omega )}{g}\right) \Gamma \left(\frac{1}{4}+\frac{i (k +\Omega )}{g}\right)}{\Gamma \left(\frac{1}{4}\right)} .
     \label{Eq:4.4.0.16}
 \end{equation}
Now, the particle number can be evaluated as,
\begin{equation}
    \begin{split}
		N(\Omega) &= \int_{0}^{\infty} d k \; \beta_{21u}^*(\Omega,k) \beta_{21u}(\Omega,k).
	\end{split}
    \label{Eq:4.4.0.17}
\end{equation}
Substituting the value of $\beta_{21u}^*(\Omega,k)$ and its complex conjugate $\beta_{21u}(\Omega,k)$ in the Eq.~(\ref{Eq:4.4.0.17}) the integrand (denoted by $G(\Omega,g,k)$), can be written as,
\begin{equation}
    G(\Omega,g,k) = \frac{|\Gamma \left(-\frac{i (k+\Omega )}{g}\right)|^2 | \Gamma \left(\frac{1}{4}-\frac{i (k+\Omega )}{g}\right)|^2}{4 \pi ^2 g^2 \Gamma \left(\frac{1}{4}\right)^2}.
    \label{Eq:4.4.0.18}
\end{equation}
From the identity $|\Gamma(x+i y)|\leq |\Gamma(x)|$ (see Ref.~\cite{Frank_Olver_822801}), we can rewrite the above as (evaluating the maximum value of the above integral),
\begin{equation}
    G(\Omega,g,k) \approx \frac{|\Gamma \left(-\frac{i (k+\Omega )}{g}\right)|^2 | \Gamma \left(\frac{1}{4}\right)|^2}{4 \pi ^2 g^2 \Gamma \left(\frac{1}{4}\right)^2}.
    \label{Eq:4.4.0.19}
\end{equation}
We now prove using a couple of justifications that the above expressions yield a finite number of particles and hence negligible number density when divided by the infinite volume. Using this as the new integrand for the particle number integral can be simplified to get,
\begin{equation}
	N(\Omega) \approx \int_{0}^{\infty} d k \; \frac{1}{4\pi g (k +\Omega )} \;\frac{1}{\sinh{\big(\frac{\pi (k +\Omega )}{g}\big)}}.
    \label{Eq:4.4.0.20}
\end{equation}
Since $k$ is always positive, we know $k+\Omega$ is always greater than $\Omega$. Hence, from these facts, the Integral in Eq.~(\ref{Eq:4.4.0.20}) can be expressed as,
\begin{equation}
   \begin{split}
       \int_{0}^{\infty} d k \;& \frac{1}{4\pi g \;(k +\Omega )\;\sinh{\big(\frac{\pi (k +\Omega )}{g}\big)}} \\
       &\leq  \int_{0}^{\infty} d k \; \frac{1}{4\pi g \;\Omega \;\sinh{\big(\frac{\pi (k +\Omega) }{g}\big)}} = N_{upp}(\Omega),
   \end{split}
   \label{Eq:4.4.0.21}
\end{equation}
this integral in the RHS gives an upper bound ($N_{upp}(\Omega)$) for our required particle number integral. By evaluating the Integral in the RHS of the Eq.~(\ref{Eq:4.4.0.21}), we get this upper bound as,
\begin{equation}
    N_{upp}(\Omega) =  \frac{\tanh ^{-1}\left(e^{\frac{\pi  \Omega }{g}}\right)}{2 \pi ^2 \Omega },
    \label{Eq:4.4.0.22}
\end{equation}
which is finite. We note that the above expression $N_{upp}(\Omega)$ is the total particle number with frequency $\Omega$ in $R_2$, and we got it as finite. To evaluate the number density $\tilde{n}(\Omega)$, we divide the total number $N_{upp}(\Omega)$ by the volume of $R_2$ given by $V$, which is infinite. This number density $N_{upp}(\Omega)$ is, therefore, negligible. Now, since $N_{upp}(\Omega)$ is always greater than $N(\Omega)$, and since from Eq.~(\ref{Eq:4.4.0.22}), $N_{upp}(\Omega)$ is finite, we can claim the Integral $N(\Omega)$ is also finite. Thus it implies that the number density $n(\Omega)$ is negligible due to the infinite spatial volume of $R_{2}$.

\subsection{Excitations of Left handed modes:\label{Subsec-4.5}}
Similarly, we perform the same exercise as in the previous section to study the excitations of the left-moving modes. We consider the mode expansion of the field operator (Majorana) in terms of the left-moving modes in the common patch of $R_1$ and $R_2$. Here also, we use the same arguments as in the previous discussion. The modes of $R_2$ are related to the modes of $R_1$ through a Bogoliubov transformation of the form Eq.~(\ref{Eq:4.4.0.1}) and Eq.~(\ref{Eq:4.4.0.3}), such that the creation/annihilation operators of $R_2$ and $R_1$ are related by Eq.~(\ref{Eq:4.4.0.4}) and Eq.~(\ref{Eq:4.4.0.5}). Thus, in the common patch of $R_1$ and $R_2$, the field operator of the left moving sector is expanded using the left modes as,
\begin{equation}
		\begin{split}
			&\int_0^\infty d k\frac{e^{\frac{-g x_1}{2}}}{\sqrt{2\pi}}\bigg(c_{1}(k)\begin{bmatrix}0\\
				e^{-ik(t_1+x_1)}
			\end{bmatrix}+c_{1}^\dagger(k)\begin{bmatrix}0\\
				e^{ik(t_1+x_1)}
			\end{bmatrix}\bigg)\\
			&=\int_0^\infty d p \frac{e^{\frac{-g x_2}{2}}}{\sqrt{2\pi}}\bigg(c_{2}(p)\begin{bmatrix}0\\
				e^{-ip(t_2+x_2)}
			\end{bmatrix}+c_{2}^\dagger(p)\begin{bmatrix}0\\
				e^{i p(t_2+x_2)}
			\end{bmatrix}\bigg).
		\end{split}
    \label{Eq:4.5.0.1}
\end{equation}
 Rewriting the above in terms of null coordinates, 
 \begin{equation}
		\begin{split}
			&\int_0^\infty dk\frac{e^{\frac{-g }{4}(v_1-u_1)}}{\sqrt{2\pi}}\bigg(c_{1}(k)\begin{bmatrix}0\\
				e^{-i k v_1}
			\end{bmatrix}+c_{1}^\dagger(k)\begin{bmatrix}0\\
				e^{i k v_1}
			\end{bmatrix}\bigg)\\
			&=\int_0^\infty dp \frac{e^{\frac{-g }{4}(v_2-u_2)}}{\sqrt{2\pi}}\bigg(c_{2}(p)\begin{bmatrix}0\\
				e^{-i p v_2}
			\end{bmatrix}+c_{2}^\dagger(p)\begin{bmatrix}0\\
				e^{i p v_2}
			\end{bmatrix}\bigg).
		\end{split}
    \label{Eq:4.5.0.2}
\end{equation}
 Now, to get the expressions for the Bogoliubov coefficients, we take the Fourier transform on both sides, i.e., we multiply both the LHS and RHS of the Eq.~(\ref{Eq:4.5.0.2}) with
 \begin{equation}
     \int_{-\infty}^{\infty} \frac{dv_2}{2 \pi} e^{\frac{g }{4}(v_2-u_2)} 
     \begin{bmatrix}
     0 & e^{i \Omega v_2}
	\end{bmatrix}.
 \label{Eq:4.5.0.3}
 \end{equation}
Thus, we get the RHS as,
\begin{equation}
	\begin{split}
	    \int\limits_0^\infty dp \bigg(c_{2}(p)\int_{-\infty}^{\infty}&\frac{dv_2}{2\pi}e^{i(\Omega - p)v_2}\\
        &+c_{2}^\dagger(p)\int_{-\infty}^{\infty}\frac{dv_2}{2\pi}e^{i(\Omega + p)v_2}\bigg),
	\end{split}
    \label{Eq:4.5.0.4}
\end{equation}
and also, the LHS as,
\begin{equation}
	\begin{split}
		\int_0^\infty dk\bigg(&c_{1}(k)\int_{-\infty}^{\infty}\frac{dv_2}{2\pi}e^{\frac{g}{4}(v_2-u_2-v_1+u_1)}e^{i(\Omega v_2 - k v_1)}\\
        &+c_{1}^\dagger(k)\int_{-\infty}^{\infty}\frac{dv_2}{2\pi}e^{\frac{g}{4}(v_2-u_2-v_1+u_1)}e^{i(\Omega v_2 + k v_1)}\bigg),
	\end{split}
    \label{Eq:4.5.0.5}
\end{equation}
The RHS can be simplified to get,
\begin{equation}
	\begin{split}
		\int_0^\infty dp\bigg(c_{2}(p)\delta(\Omega-p)+c_{2}^\dagger(p)\delta(\Omega+p)\bigg)=c_{2}(\Omega).
	\end{split}
    \label{Eq:4.5.0.6}
\end{equation}
 From the Bogoliubov transformations [see Eq~(\ref{Eq:4.4.0.1}) and Eq~(\ref{Eq:4.4.0.2})], the operators of $R_1$ and $R_2$ are related by Eq~(\ref{Eq:4.4.0.3}) and Eq~(\ref{Eq:4.4.0.4}), thus from these relations and the Eq~(\ref{Eq:4.5.0.6}) and Eq~(\ref{Eq:4.5.0.5}), the Bogoliubov coefficients can be written as,
 \begin{equation}
     \alpha_{21v}^*(\Omega,k)=\int_{-\infty}^\infty\frac{dv_2}{2\pi}e^{\frac{g}{4}(v_2-u_2-v_1+u_1)}e^{i(\Omega v_2 - k v_1)},
     \label{Eq:4.5.0.7}
 \end{equation}
 and 
 \begin{equation}
     \beta_{21v}^*(\Omega,k)=\int_{-\infty}^\infty\frac{dv_2}{2\pi}e^{\frac{g}{4}(v_2-u_2-v_1+u_1)}e^{i(\Omega v_2 + k v_1)}.
     \label{Eq:4.5.0.8}
 \end{equation}
 To study the particle spectrum in $R_2$, we must evaluate the Bogoliubov coefficient $\beta_{21v}^*(\Omega,k)$. For that, on the common patch of $R_1$ and $R_2$, we will consider a constant time hypersurface $t_2 =0$, such that from Eq.~(\ref{Eq:2.1.1.8.a}) and Eq.~(\ref{Eq:2.1.1.8.b}) we get $u_2=-v_2$ . Also, as per the scenario described by the Fig.~\ref{Fig:1} and from Eq~(\ref{Eq:3.2.0.6}), we get $u_1 = u_2$. Using all these facts, and Eq~(\ref{Eq:3.3.0.7}), the integral for the Bogoliubov coefficient $\beta_{21v}^*(\Omega,k)$ can be simplified as,
 \begin{equation}
     \beta_{21v}^*(\Omega,k)=\int_{-\infty}^\infty\frac{dv_2}{2\pi}e^{\frac{g}{4}(v_2-\frac{ln{[e^{g v_2}+\lambda]}}{g})}e^{i(\Omega v_2 + k \frac{ln{[e^{g v_2}+\lambda]}}{g})}.
     \label{Eq:4.5.0.9}
 \end{equation}
Substituting $h=e^{gv_2}$, such that $dv_2 = \frac{dh}{gh}$ and applying the new limits for the integration, we get,
 \begin{equation}
     \beta_{21v}^*(\Omega,k)=\frac{1}{2\pi g}\int_{0}^\infty dh \bigg[ \frac{h}{h+\lambda} \bigg]^\frac{1}{4} h^{\frac{i \Omega}{g}-1} \big( h+\lambda \big)^{\frac{i k}{g}}.
     \label{Eq:4.5.0.10}
 \end{equation}
This integral can be evaluated to get,
\begin{equation}
     \beta_{21v}^*(\Omega,k)=\frac{\lambda ^{\frac{i (k+\Omega )}{g}}}{2\pi g} \frac{\Gamma \left(\frac{i \Omega }{g}+\frac{1}{4}\right) \Gamma \left(-\frac{i (k+\Omega )}{g}\right)}{\Gamma \left(\frac{1}{4}-\frac{i k}{g}\right)}.
     \label{Eq:4.5.0.11}
 \end{equation}
Also, its complex conjugate is,
\begin{equation}
     \beta_{21v}(\Omega,k)=\frac{\lambda ^{\frac{-i (k+\Omega )}{g}}}{2\pi g} \frac{\Gamma \left(\frac{-i \Omega }{g}+\frac{1}{4}\right) \Gamma \left(\frac{i (k+\Omega )}{g}\right)}{\Gamma \left(\frac{1}{4}+\frac{i k}{g}\right)}.
     \label{Eq:4.5.0.12}
 \end{equation}
Now, in order to find the particle number density, we first compute the particle number,
\begin{equation}
    \begin{split}
		N(\Omega) &= \int_{0}^{\infty} d k \; \beta_{21v}^*(\Omega,k) \beta_{21v}(\Omega,k),
	\end{split}
    \label{Eq:4.5.0.13}
\end{equation}
and then divide this with the volume of $R_2$. Substituting the value of $\beta_{21v}^*(\Omega,k)$ and $\beta_{21v}(\Omega,k)$ in the above expression, we get particle number integral. The integrand $|\beta_{21v}(\Omega,k)|^2$ of this integral can be expanded as,
\begin{equation}
    \begin{split}
        |\beta_{21v}(\Omega,k)|^2 &=  \frac{|\Gamma \left(\frac{-i \Omega }{g}+\frac{1}{4}\right)|^2 |\Gamma \left(\frac{i (k+\Omega )}{g}\right)|^2}{ 4 \pi^2 g^2 |\Gamma \left(\frac{1}{4}+\frac{i k}{g}\right)|^2}.
    \end{split}
    \label{Eq:4.5.0.14}
\end{equation}
Using some of the important properties of gamma functions, like the reflection formula and the asymptotic expansion formula \cite{abramowitz1968handbook, Frank_Olver_822801}, we can approximate the integrand and simplify it. After evaluating the particle number integral [see Appendix.~\ref{Apn.1} for more details], we get,
\begin{equation}
    N(\Omega) \sim \text{sech}\left(\frac{2 \pi  \Omega }{g}\right) \int_{0}^{\infty} d k \bigg[ \frac{1}{ 4 \:\pi \:g\: \sqrt{\Omega}\:\sqrt{k} } \bigg].
    \label{Eq:4.5.0.15}
\end{equation}
Here, in the above expression, the integral diverges, and from dimensional analysis, we can conclude that this divergent part has spatial units. Though the integrand is a result of an asymptotic form of the Gamma functions, the formula is continued to lower values of $k$. At lower values of k, the integrand is finitely bound. The integral evaluates to infinity only when considering larger values of $k$.  Thus, when we evaluate the number density from this particle number by dividing it with the infinite spatial volume of $R_2$. Here, we assume that dividing by the infinite volume cancels the divergent term, resulting in a finite number density distribution.  Thus we can evaluate the number density distribution for large $\Omega$ as,
\begin{equation}
   n(\Omega) \sim \frac{1}{e^{\frac{2 \pi  \Omega }{g}}}. 
    \label{Eq:4.5.0.16}
\end{equation}
We note that this expression is valid only for large $\Omega$. This matches with a thermal distribution of fermions with a temperature of $g/2\pi$ at large frequency. The standard thermal distribution for fermions matches with this expression at large frequencies $1/ (e^{\frac{2 \pi  \Omega }{g}}+1) $ since we can ignore the $1$ in the denominator at large frequencies $\Omega$. So, we may interpret this as thermal distribution, at least at large frequencies. 
This particle number density distribution and the corresponding chiral excitations for the left-moving modes are much more than what we got in the case of Right-moving modes.

\section{ Conclusions and Discussion \label{Sec-5}}
The work done in the article shows that the particle density observed in Rindler spacetime can have quantum hairs.  If one observes particle density of only left-moving or right-moving modes (or for the case of massless fermions left-handed or right-handed fermions),  it is possible to deduce the causal placement of the larger spacetime, whose vacuum state gives rise to the observed particle density (purifications of the Rindler wedge).  The calculation done in this article is in the context of two-dimensional Rindler spacetime and quantum fields living in it. Though the work is in a more elementary setting, a few obvious questions can be raised.  Is there selective excitation possible when one generalizes the massive fields also? For massive particles, the left-moving modes and right-moving modes become frame-dependent.  Are there situations where chiral asymmetric excitation is possible, even for massive Dirac fields? Are the results of this article valid for four dimensions also? These are some of the more obvious questions that can be asked based on the results of the paper.
\par A slightly more challenging set of questions yet to be investigated may be asked along the work done in this article. How does one solve for the entanglement entropy across a nested sequence of Rindler observers shifted from each other by spacelike or null intervals via information-theoretic techniques or the techniques of algebraic quantum field theory? \cite{witten1,Witten.045003, headrik,Nishioka.035007}. A standard method used to derive the particle content in Rindler spacetime from a Minkowski vacuum state is via the method of Euclidean path integral formulation proposed in the celebrated paper by Gibbons and Hawking \cite{Gibbonsandhawking}.  How does one extend the technique to solve for the particle density across a sequence of Rindler spacetimes shifted along spacelike or null intervals?
\par As pointed out earlier in the draft, the shifted Rindler spacetimes or the nested sequence of Rindler spacetimes may offer important clues towards capturing quantum signatures from evolving horizons. During the radiation-dominated era of cosmological evolution, the marginally trapped region evolves along a light-like direction \cite{bendov,raviteja}. One can represent this evolution by a series of wedges shifted along a null direction. If one considers massless scalar field or Dirac field in this spacetime, is it possible that in this cosmological era, only left-handed or right-handed particles get excited, thereby creating asymmetry in the particle density in terms of handedness for massless fermions or low mass fermions like neutrinos? These questions are left for later considerations.
\par In this article, we could show that in the case of massless scalar fields, the left-moving modes are thermally excited, whereas the right-moving modes are in vacuum. In contrast, in the case of massless fermionic fields, the right-handed modes have negligible particle density, whereas the left-handed modes are excited and have non-zero particle density. The spectrum, whether it is thermal or not for all modes, is unclear at this stage. It is thermal for high-frequency modes. The result indicates a quantum of hair in Rindler spacetime. If one detects selective thermalization of only the left moving modes, then Fig.~\ref{Fig:1} is the relevant setting. If, on the other hand, one detects selective thermalization of only right-moving modes, then Fig.~\ref{Fig:2} is the relevant setting. A similar conclusion is reached if one looks at massless fermionic fields based on whether right-handed or left-handed particles are excited. So, the asymmetric excitations will let one identify the placement of the superset of a given Rindler spacetime that yields the observed particle density. The information of the causal placement of the super set is therefore decipherable from the observed particle density. This shows that informative excitement is possible in Rindler spacetime. The Rindler spacetimes can, therefore, have quantum hairs. Whether there is any implication of the above results on the information loss problem is yet to be explored.

\appendix
\section{Particle Number density for the Left-moving modes: \label{Apn.1}}
To find the particle number density for the left moving modes, we first compute the particle number,
\begin{equation}
    \begin{split}
		N(\Omega) &= \int_{0}^{\infty} d k \; \beta_{21v}^*(\Omega,k) \beta_{21v}(\Omega,k),
	\end{split}
\end{equation}
and then divide this with the volume of $R_2$. 
Substituting the value of $\beta_{21v}^*(\Omega,k)$ and $\beta_{21v}(\Omega,k)$ from Eq.~(\ref{Eq:4.5.0.11}) and Eq.~(\ref{Eq:4.5.0.12}), in the above expression, we get particle number integral. The integrand of this integral can be denoted as,
\begin{equation}
    \begin{split}
        |\beta_{21v}(\Omega,k)|^2 &=  \frac{|\Gamma \left(\frac{1}{4} -\frac{i \Omega }{g}\right)|^2 |\Gamma \left(\frac{i (k+\Omega )}{g}\right)|^2}{ 4 \pi^2 g^2 |\Gamma \left(\frac{1}{4}+\frac{i k}{g}\right)|^2}.
    \end{split}
    \label{Apn.A2}
\end{equation}
To simplify the integrand, we use the well-known results for the gamma functions, the Reflection formula \cite{abramowitz1968handbook}, which gives
\begin{equation}
    \Gamma{(1-z)} = \frac{\pi}{\Gamma{(z)} \sin{\pi z}},
\end{equation}
and the asymptotic expansion for the Gamma functions \cite{Frank_Olver_822801}, which gives,
\begin{equation}
    |\Gamma{(x+i y)}| \sim \sqrt{2 \pi} |y|^{x-\frac{1}{2}} e^{-\frac{\pi |y|}{2}}.
    \label{Apn.A4}
\end{equation}
First, consider the numerator, the function $\Gamma \left(\frac{1}{4} -\frac{i \Omega }{g}\right)$ can be replaced by,
\begin{equation}
    \Gamma \left(\frac{1}{4} -\frac{i \Omega }{g}\right)=\frac{\pi }{\sin \left(\pi  \left(\frac{3}{4}+\frac{i \Omega }{g}\right)\right) \Gamma \left(\frac{3}{4}+\frac{i \Omega }{g}\right)}.
\end{equation}
Therefore we get,
\begin{equation}
    |\Gamma \left(\frac{1}{4} -\frac{i \Omega }{g}\right)|^2 = \frac{2 \pi ^2 \;\text{sech}\left(\frac{2 \pi  \Omega }{g}\right)}{\Gamma \left(\frac{3}{4}-\frac{i \Omega }{g}\right) \Gamma \left(\frac{3}{4}+\frac{i \Omega }{g}\right)}.
    \label{Apn.A6}
\end{equation}
Now using the Eq.~(\ref{Apn.A4}) (asymptotic expansion of Gamma function), the $\Gamma \left(\frac{3}{4}-\frac{i \Omega }{g}\right)$ becomes,
\begin{equation}
    \Gamma \left(\frac{3}{4}-\frac{i \Omega }{g}\right) \sim \sqrt{2 \pi}  \left(\frac{\Omega}{g}\right)^{\frac{3}{4}-\frac{1}{2}} e^{-\frac{\pi  \Omega}{2\: g}},
\end{equation}
and the denominator of Eq.~(\ref{Apn.A6}) becomes,
\begin{equation}
    |\Gamma \left(\frac{3}{4}-\frac{i \Omega }{g}\right)|^2 \sim 2 \pi  \left(\frac{\Omega}{g}\right)^{\frac{1}{2}} e^{-\frac{\pi  \Omega}{g}}.
\end{equation}
Thus Eq.~(\ref{Apn.A6}) can be rewritten as,
\begin{equation}
    |\Gamma \left(\frac{1}{4} -\frac{i \Omega }{g}\right)|^2 = \frac{2 \pi ^2 \:\sqrt{g}\;\text{sech}\left(\frac{2 \pi  \Omega }{g}\right)}{2 \:\pi\: \sqrt{\Omega} \:e^{-\frac{\pi  \Omega}{g} }}.
    \label{Apn.A9}
\end{equation}
The other term in the numerator is $|\Gamma \left(\frac{i (k+\Omega )}{g}\right)|^2$, it when simplified results in,
\begin{equation}
    |\Gamma \left(\frac{i (k+\Omega )}{g}\right)|^2 = \frac{\pi  g \;\text{csch}\left(\frac{\pi  (k+\Omega )}{g}\right)}{(k+\Omega)}.
    \label{Apn.A10}
\end{equation}
Now considering the denominator, we use the asymptotic expression to get,
\begin{equation}
    |\Gamma \left(\frac{1}{4}+\frac{i k}{g}\right)| \sim \sqrt{2 \pi } \left(\frac{k}{g}\right)^{\frac{1}{4}-\frac{1}{2}} e^{-\frac{\pi  k}{ 2\:g}},
\end{equation}
therefore we get,
\begin{equation}
    |\Gamma \left(\frac{1}{4}+\frac{i k}{g}\right)|^2 \sim 2 \pi  \left(\frac{k}{g}\right)^{-\frac{1}{2}} e^{-\frac{\pi  k}{ g}}.
    \label{Apn.A12}
\end{equation}
Now substituting Eq.~(\ref{Apn.A9}), Eq.~(\ref{Apn.A10}) and Eq.~(\ref{Apn.A12}) in Eq.~(\ref{Apn.A2}), the integrand expression gives,
\begin{equation}
    |\beta_{21v}(\Omega,k)|^2=\frac{\text{sech}\left(\frac{2 \pi  \Omega }{g}\right)}{ e^{-\frac{\pi \Omega}{g}}}\frac{ \sqrt{k}\;\text{csch}\left(\frac{\pi  (k+\Omega )}{g}\right)}{ 8 \:\pi \:g\: \sqrt{\Omega}\:(k+\Omega ) e^{-\frac{\pi  k}{ g}} }.
\end{equation}
Thus the particle number integral can now be written as,
\begin{equation}
    N(\Omega) = F(\Omega) \int_{0}^{\infty} d k \bigg[\frac{ \sqrt{k}\;\text{csch}\left(\frac{\pi  (k+\Omega )}{g}\right)}{ 8 \:\pi \:g\: \sqrt{\Omega}\:(k+\Omega ) e^{-\frac{\pi  k}{ g}} } \bigg], 
\end{equation}
where $F(\Omega)$ is the form of particle spectrum in terms of $\Omega$.
Now, to perform the integral, we consider the asymptotic form of the function at large $k$. Therefore for $k\rightarrow \infty$, we can approximate the $\text{csch}\left(\frac{\pi  (k+\Omega )}{g}\right) \sim 2 \;e^{- \frac{\pi(k+\Omega)}{g}}$ and for  $(k+\Omega) \sim k$. Thus we get,
\begin{equation}
    N(\Omega) \sim F(\Omega)\:e^{-\frac{\pi \Omega}{g}} \int_{0}^{\infty} d k \bigg[\frac{ \sqrt{k}\;2 \;e^{- \frac{\pi(k+\Omega)}{g}}}{ 8 \:\pi \:g\: \sqrt{\Omega}\:k \:e^{-\frac{\pi  k}{ g}} }\bigg], 
\end{equation}
and on further simplifying, we get,
\begin{equation}
    N(\Omega) \sim \text{sech}\left(\frac{2 \pi  \Omega }{g}\right) \int_{0}^{\infty} d k \bigg[ \frac{1}{ 4 \:\pi \:g\: \sqrt{\Omega}\:\sqrt{k} } \bigg]. 
\end{equation}
Converting the above into exponential terms, we get,
\begin{equation}
    N(\Omega) \sim \bigg[\frac{2 e^{\frac{2 \pi  \Omega }{g}}}{e^{\frac{4 \pi  \Omega }{g}}+1} \bigg]\int_{0}^{\infty} d k \bigg[ \frac{1}{ 4 \:\pi \:g\: \sqrt{\Omega}\:\sqrt{k} } \bigg]. 
\end{equation}
The above integral diverges now when we evaluate the number density from the particle number by dividing the above with an infinite volume of $R_2$ we get,
\begin{equation}
   n(\Omega) \sim \frac{e^{\frac{2\pi  \Omega }{g}}}{e^{\frac{4 \pi  \Omega }{g}}+1}. 
\end{equation}
This number density, when considered for large $\Omega$, can be written as,
\begin{equation}
   n(\Omega) \sim \frac{1}{e^{\frac{2 \pi  \Omega }{g}}}. 
\end{equation}
Thus, we get a significant particle number density for the left-moving modes.
\begin{acknowledgments}
We thank our institute, BITS Pilani Hyderabad campus, for providing the required infrastructure for this research work.
\end{acknowledgments}

\bibliography{Ref.bib}

\end{document}